\documentclass[notitlepage,prd,showpacs,preprintnumbers,nofootinbib,superscriptaddress,floatfix,tightenlines, longbibliography]{revtex4-1}
\usepackage[utf8]{inputenc}
\usepackage{bm}
\usepackage{mathtools}
\usepackage{amsmath}
\usepackage{amsfonts}
\usepackage[toc,page]{appendix}
\usepackage{dsfont}
\usepackage{comment}
\usepackage{bbold}
\usepackage{soul}
\usepackage{nicefrac}
\usepackage[dvipsnames]{xcolor}
\usepackage[colorlinks=true,linkcolor=blue,citecolor=blue, urlcolor=blue]{hyperref}  
\usepackage[a4paper,top=2.4cm,bottom=2.4cm,right=1.8cm,bindingoffset=-2cm]{geometry}

\usepackage{tcolorbox}

\usepackage{stackengine,xcolor}
\usepackage{appendix}

\newcommand{\fig}{Fig.~}

\newcommand{\eq}{Eq.~}
\newcommand{\eqs}{Eqs.~}
\newcommand{\se}{Sec.~}

\renewcommand{\rm}[1]{\mathrm{#1}}

\usepackage{xargs}
\usepackage[colorinlistoftodos,prependcaption,textsize=footnotesize]{todonotes}

\newcommandx{\KB}[2][1=]{\todo[linecolor=orange,backgroundcolor=orange!25,bordercolor=orange,#1]{KB: #2}}

\newcommandx{\TL}[2][1=]{\todo[linecolor=purple,backgroundcolor=purple!25,bordercolor=purple,#1]{TL: #2}}

\newcommandx{\JP}[2][1=]{\todo[linecolor=blue,backgroundcolor=blue!25,bordercolor=blue,#1]{JP: #2}}

\newcommandx{\PS}[2][1=]{\todo[linecolor=red,backgroundcolor=red!25,bordercolor=red,#1]{PS: #2}}

\newcommandx{\TODO}[2][1=]{\todo[linecolor=red,backgroundcolor=red!25,bordercolor=red,#1]{TODO: #2}}
\begin{document}

\title{Conserved energy-momentum tensor for real-time lattice simulations}

\author{K.~Boguslavski} 
\affiliation{Institute for Theoretical Physics, Technische Universit\"{a}t Wien, 1040 Vienna, Austria}

\author{T.~Lappi} 
\affiliation{Department of Physics, University of Jyv\"{a}skyl\"{a}, P.O.~Box 35, 40014 University of Jyv\"{a}skyl\"{a}, Finland}
\affiliation{Helsinki Institute of Physics, P.O.~Box 64, 00014 University of Helsinki, Finland}

\author{J.~Peuron} 
\affiliation{Department of Physics, University of Jyv\"{a}skyl\"{a}, P.O.~Box 35, 40014 University of Jyv\"{a}skyl\"{a}, Finland}
\affiliation{Helsinki Institute of Physics, P.O.~Box 64, 00014 University of Helsinki, Finland}

\author{P.~Singh} 
\email{pragya.phy.singh@jyu.fi}
\affiliation{Department of Physics, University of Jyv\"{a}skyl\"{a}, P.O.~Box 35, 40014 University of Jyv\"{a}skyl\"{a}, Finland}
\affiliation{Helsinki Institute of Physics, P.O.~Box 64, 00014 University of Helsinki, Finland}

\begin{abstract}
We derive an expression for the energy-momentum tensor in the discrete lattice formulation of pure glue QCD. The resulting expression satisfies the continuity equation for energy conservation up to numerical errors with a symmetric procedure for the time discretization. In the case of the momentum conservation equation, we obtain an expression that is of higher accuracy in lattice spacing ($\mathcal{O}(a^2)$) than the naive discretization where fields in the continuum expressions are replaced by discretized counterparts. The improvements are verified by performing numerical tests on the derived expressions using classical real-time lattice gauge theory simulations. We demonstrate substantial reductions in relative error of one to several orders of magnitude compared to a naive discretization for both energy and momentum conservation equations. We expect our formulation to have applications in the area of pre-equilibrium dynamics in ultrarelativistic heavy ion collisions, in particular for the extraction of transport coefficients such as shear viscosity. 
\end{abstract}

\maketitle

\section{Introduction}
Lattice gauge theory \cite{Wilson:1974sk} is a powerful tool commonly used to address nonperturbative problems in Quantum Chromodynamics (QCD). This lattice framework encodes gauge invariance by construction and therefore preserves this crucial property of gauge theory. However, the lattice discretization modifies or breaks some of the underlying symmetries of the theory, such as translational or rotational invariance. 

The energy-momentum tensor (EMT) is an observable that encodes energy and momentum conservation in the form of a continuity equation. The conserved quantities in this case arise from invariance under space-time translations as described by Noether's theorem. Furthermore, the energy-momentum tensor is traceless at the classical level due to conformal symmetry. In a lattice discretized formulation, these symmetries are modified, as invariance under space-time translations becomes a discrete symmetry instead of a continuous one. The conformality of the theory is also broken by involving an explicit scale, the lattice spacing. 
In spite of these issues, one typically evaluates the classical energy-momentum tensor by taking its continuum expression and replacing the fields with their discretized counterparts. It is not clear a priori that the resulting expression satisfies the continuity equation. 

Our interest in this topic is motivated by the early-time dynamics in the context of ultrarelativistic heavy-ion collisions. Directly after the collision the system consists of over-occupied gluon fields, is typically modeled using classical Yang-Mills theory, and is often referred to as the Glasma \cite{Gelis:2010nm, Gelis:2012ri, Lappi:2006fp}.
Recently, the Glasma stage has received substantial attention \cite{Schenke:2012wb, Schenke:2012hg, Lappi:2017skr, Albacete:2018bbv, Fujii:2008km, Berges:2013eia, Berges:2013fga, Boguslavski:2019fsb, Carrington:2021qvi, Carrington:2020ssh, Ipp:2017lho, McDonald:2023qwc, Schenke:2016ksl, Schlichting:2020wrv, Ipp:2021lwz, Schenke:2022mjv, Matsuda:2023gle}, especially in terms of its properties as a medium, described by transport coefficients \cite{Boguslavski:2020tqz, Avramescu:2023qvv, Ipp:2020nfu, Carrington:2020sww, Pooja:2023ubs, Pooja:2022ojj, Carrington:2022bnv, Carrington:2021dvw}. 

This paper aims to construct an improved expression for the lattice energy-momentum tensor of a nonabelian gauge field following classical equations of motion, such that the tensor satisfies the continuity equation exactly whenever possible and improves the accuracy of energy-momentum conservation when this is not the case. As our applications lie in the domain of real-time simulations \cite{Lappi:2003bi, Lappi:2006fp, Schenke:2012wb, Schenke:2012hg, Berges:2013eia, Berges:2013fga, Boguslavski:2019fsb, Ipp:2017lho, McDonald:2023qwc, Schenke:2016ksl, Schlichting:2020wrv, Ipp:2021lwz, Schenke:2022mjv, Avramescu:2023qvv, Boguslavski:2020tqz, Ipp:2020nfu, Shen:2022oyg}, where a classical-statistical approximation is employed, our approach will inevitably differ from that used in lattice QCD. This is due to the fact that the EMT in lattice QCD is constructed to satisfy Ward-Takahashi identities \cite{Caracciolo:1989pt, Caracciolo:1991cp, Bochicchio:1985xa} at the quantum level after renormalization. These identities play the role of Noether's theorem for quantum field theories. Hence, while our approach is similar aiming to capture the same physical phenomena, the two approaches are nevertheless not identical.

For an impatient reader, who is only interested in our main results and less in the technical details concerning their derivation, we summarize our results as follows. The energy density component $T^{00}$ of the EMT is given by \eq \eqref{eq:T00_basic}, which redistributes the total energy on the lattice into local energy densities that are spatially symmetric and centered at a lattice site. The Poynting vector components $T^{0i}$ are then obtained by requiring that a discretized continuity equation for energy conservation be satisfied. 
This procedure 
 leads to $T^{0i}$ given by \eq \eqref{eq:T_{0k}} that satisfies the energy conservation continuity equation up to numerical errors. We will also discuss how to further improve this expression by making use of time symmetrization to synchronize the electric and magnetic fields in \se \ref{sec:symm_time_discr}.
For the spatial components $T^{ij}$, we start from the Poynting vector and impose the continuity equation for momentum conservation. This leads to an expression for the discrete analog of the Maxwell stress tensor $T^{ij}$. However, within this derivation, we will have to perform a few approximations due to additional complications. These arise from parallel transport on the lattice and from the fact that on the lattice we do not have a suitable analog of the continuum Bianchi identity. Our final result is given by \eq \eqref{eq:final_Tjk} and, as we will show in a separate section, it satisfies the continuity equation to an accuracy of order $\mathcal{O}(a_s^2)$. 

We believe that the idea and the algorithm presented here to derive a conserved $T_{\mu\nu}$ (with high precision) can be utilized much more broadly. Nonconservation of $T^{\mu \nu}$ is expected to occur in every theory discretized on a lattice due to the breaking of the space-time translational symmetry. Particularly interesting examples of potential applications beyond heavy-ion collisions are real-time lattice simulations performed in the context of postinflatory cosmology using the classical approximation \cite{Tranberg:2022noe, Arrizabalaga:2004iw, DOnofrio:2014rug}. 

This paper is structured as follows. \se \ref{se:EOMSandFormalism} describes our discretization framework. In \se \ref{se:continuity} we construct the discrete energy-momentum tensor and introduce its time-symmetrized formulation. \se \ref{se:numerics} shows the numerical comparison among the naive and discretized approaches. We conclude in \se \ref{se:conclu}.

\section{General formalism and classical equations of motion}
\label{se:EOMSandFormalism}

In line with established practice within classical-statistical lattice gauge theory simulations, we begin with the discrete Kogut-Susskind Hamiltonian \cite{Kogut:1974ag} of classical SU($N_c$) Yang-Mills theory in temporal axial gauge ($A_t = 0$),
\begin{align}
    \label{eq:CYMhamilton:orig}
    \mathcal{H} &= \sum_{m}a_s^3\Bigg(-\frac{1}{2\sqrt{-g}}\sum_{\mu>0}g_{\mu\mu}E^{\mu,a}_mE^{\mu,a}_m + \frac{2N_c}{g^2} \sum_{0<\mu<\nu}\frac{1}{a_\mu^2a_\nu^2}g^{\mu\mu}g^{\nu\nu}\sqrt{-g}\Big(1-\frac{1}{N_c}{\rm{ReTr}}(U_{\mu\nu,m})\Big)\Bigg) \\
    &= \sum_{m}a_s^3\Bigg(\frac{1}{2}\sum_{\mu>0} E^{\mu,a}_mE^{\mu,a}_m + \frac{2N_c}{g^2} \sum_{0<\mu<\nu}\frac{1}{a_\mu^2a_\nu^2}\Big(1-\frac{1}{N_c}{\rm{ReTr}}(U_{\mu\nu,m})\Big)\Bigg).
    \label{eq:CYMhamilton}
\end{align}
Here $m$ denotes the lattice site on the spatial grid and $a=1, \ldots ,N_c^2-1$ is the color index. In this paper, we consider the Minkowski metric with $g_{\mu\nu}=(+1,-1,-1,-1)$ and $\sqrt{-g} = \sqrt{-\textrm{det}(g_{\mu\nu})} = 1$. The coupling constant $g$ enters in the form of the gauge coupling $2N_c/g^2$.
The discretization is performed on a three-dimensional lattice with size $N_x \times N_y \times N_z$, lattice spacing $a_\mu$ in the spatial direction $\hat{\mu}$, and the product $a_s^3 = a_x a_y a_z$. 
To guarantee gauge invariance, the theory is formulated in terms of gauge links $U_{i,m}= e^{ia_i gA_{i,m}}$ instead of gauge fields $A_{i,m}$. Links enter \eq \eqref{eq:CYMhamilton} in terms of plaquettes
\begin{equation}
U_{\mu\nu,m}=U_{\mu,m}U_{\nu,m+\hat{\mu}}U^\dagger_{\mu,m+\hat{\nu}}U^\dagger_{\nu,m}\,,
\end{equation}
where $\hat \mu$ denotes a unit vector in the $\mu$ direction.%
\footnote{To simplify the notation, we will often write $\mu$ instead of $\hat \mu$.}
In analogy to the links, plaquettes are related to the field-strength tensor $F_{\mu\nu}$ via $U_{\mu\nu} \approx e^{i g a_\mu a_\nu F_{\mu\nu}}$.

Links and plaquettes are group elements $U_{\mu,m}, U_{\mu\nu,m} \in$ SU($N_c$) in the fundamental representation, and correspondingly we will use the generators of the fundamental representation $t^a$ to go between the color components of the electric field and the fundamental representation matrix $E^{\mu}_m=E^{\mu,a}_m t^a$. Using the (continuum) relation between the field-strength tensor and the chromomagnetic field $B^i = -\frac{1}{2} \epsilon^{ijk} F_{jk}$,
the Hamiltonian \eqref{eq:CYMhamilton} reduces to the correct expression $\int d^3x \frac{1}{2}\sum_{j>0}(E^{j,a} E^{j,a} + B^{j,a} B^{j,a})$ in the continuum limit $a_\mu \to 0$.

The Hamiltonian \eqref{eq:CYMhamilton} is associated with%
\footnote{We have rescaled the chromoelectric fields $E^{\mu,a}_m$ to correspond to the continuum field, with dimension GeV$^2$, thus they differ from the canonical momentum variables of the discrete theory by a multiplicative factor.} 
the Hamiltonian equations of motion for the link matrices and the electric field variables, which are discrete in space but continuous in time
\begin{align}
    \partial_0 U_{\mu,m}&{}=iga_\mu E^{\mu,a}_m t^a U_{\mu,m}\label{eq:evol_U}\\
     \partial_0 \,g E^{\mu,a}_m&{}=\sum_{0<\nu\neq\mu}\frac{2 a_\mu}{a_\mu^2a_\nu^2}\rm{ReTr}\big(it^aU_{\mu\nu,m}+it^aU_{\mu-\nu,m}\big)\label{eq:prestep_evol_E}\\
    &{}=\sum_{0<\nu\neq\mu}\frac{2}{a_\mu a_\nu}\rm{ReTr}\big(it^a[D_\nu^B,U_{\mu\nu,m}]\big).\label{eq:evol_E}
\end{align}
These equations of motion are typically solved using a leapfrog algorithm, which is second-order accurate in the time step $dt$. In this discretization paradigm, the electric fields and links are located half a timestep apart. This is illustrated in the right panel of \fig \ref{fig:Lattice_T00}.

In order to derive Eq.~\eqref{eq:evol_E} from Eq.~\eqref{eq:prestep_evol_E}, we have used the identity $U_{\mu-\nu,m}^\dagger = U_{\nu,m-\hat\nu}^\dagger U_{\mu\nu,m-\hat\nu}U_{\nu,m-\hat\nu}$ and introduced the backward and forward gauge covariant derivatives
\begin{align}
\label{eq:defDB}
    D_{\mu,m}^BX={}&\big(X_{m}-U^\dagger_{\mu,m-\hat{\mu}}X_{m-\hat{\mu}}~U_{\mu,m-\hat{\mu}}\big)/a_\mu \\
\label{eq:defDF}
    D_{\mu,m}^FX={}&\big(U_{\mu,m}X_{m+\hat{\mu}}U^\dagger_{\mu,m}-X_{m}\big)/a_\mu.
\end{align}
By utilizing the equation of motion for the link matrices Eq.~\eqref{eq:evol_U}, one can deduce the time derivative of the magnetic field part of the energy density
\begin{align}
    \partial_0{\rm{ReTr}}\,U_{\mu\nu,m}
    &{}={\rm{ReTr}}\Big(ig a_\mu a_\nu\big[D_\mu^F,E^\nu_m\big]U_{\mu\nu,m}-iga_\mu a_\nu\big[D_\nu^F,E^\mu_m\big]U_{\mu\nu,m} \Big).
    \label{eq:d0ReTrUjk}
\end{align}

\section{Constructing the energy-momentum tensor}
\label{se:continuity}

In this section, we outline our methodology for acquiring a discretized representation of the energy-momentum tensor (EMT).
In a broad sense, Noether's theorem introduces the framework for establishing a connection between the temporal and spatial translation invariance inherent in Yang-Mills theory and the energy-momentum tensor. In the continuum, the canonical EMT can then be computed from Noether's theorem and leads to a non-symmetric tensor that requires an additional gauge transformation to become symmetric. Instead, one can use the (Hilbert) EMT $T^{\mu\nu}$ obtained by taking a functional derivative of the Yang-Mills action $S = \sum_m a_s^3 \sqrt{-g} \mathcal L$ with respect to the metric tensor $g_{\mu\nu}$,
\begin{align}
\label{eq:Tmunuvariation}
    T^{\mu\nu} = \frac{2}{\sqrt{-g}} \frac{\delta S}{\delta g_{\mu\nu}} = 2\frac{\partial \mathcal{L}}{\partial g_{\mu\nu}}-g^{\mu\nu}\mathcal{L}.
\end{align}
On the lattice, our starting point is a situation where the metric is purely diagonal. Indeed, both the Hamiltonian \eqref{eq:CYMhamilton:orig} and the corresponding action $S$ only include diagonal metric elements. Hence, there is no straightforward way to vary off-diagonal components of the metric in a continuous way around zero in order to calculate derivatives with respect to the metric as in~\eqref{eq:Tmunuvariation}. This approach is therefore not applicable to our purposes. Furthermore, off-diagonal terms play a crucial role when investigating transport coefficients, such as shear viscosity.

We will first describe in \se \ref{sec:EMT_naive} how the EMT could be constructed naively inspired by the continuum expressions. Then we will explain our improved procedure that is directly based on the continuity equation for energy-momentum conservation
\begin{align}
     \partial_\mu T^{\mu\nu}  =0.
    \label{eq:dmu_Tmunu}
\end{align}
We will start in \se \ref{sec:E_conserv} by allowing spatial redistribution for the energy density, i.e., the temporal component $T^{00}$ while requiring that it sums up to the Hamiltonian that corresponds to the total energy \eqref{eq:CYMhamilton}. The components $T^{0i}$ are then constructed from \eqref{eq:dmu_Tmunu} for $\nu = 0$. 
The remaining components will then be constructed in \se \ref{sec:T0i_conserv} in the same way by taking $T^{0i}$ as the starting point and requiring that $T^{ij}$ satisfy the remaining continuity equations for $\nu > 0$. 
This procedure determines $T^{0i}$ up to a transformation $T^{0i} \to T^{0i} + \phi^i$, where $\partial_i \phi^i = 0$. 
This transformation, however, leaves conserved quantities intact.

\subsection{Energy momentum tensor in the continuum and naive discretization}
\label{sec:EMT_naive}

The naive discretization of the energy-momentum tensor proceeds by taking the continuum expression and replacing $E$ and $B$ fields with their spatially averaged discretized counterparts
\begin{align}
    T^{00}_{m}&=\frac{1}{2}\big(E^2_{m,\rm{loc}}+B^2_{m,\rm{loc}}) \label{eq:T00_cont}\\ 
    T^{0i}_m&=\epsilon^{ijk}\big(E^j_{m,\rm{loc}} B^k_{m,\rm{loc}}\big)\label{eq:T0i_cont}\\
    T^{ij}_m&=\frac{1}{2}\delta^{ij}\big(E^2_{m,\rm{loc}} + B^2_{m,\rm{loc}}\big) - E^i_{m,\rm{loc}} E^j_{m,\rm{loc}} - B^i_{m,\rm{loc}} B^j_{m,\rm{loc}} \nonumber \\
    &= \delta^{ij}T^{00}_m - E^i_{m,\rm{loc}} E^j_{m,\rm{loc}} - B^i_{m,\rm{loc}} B^j_{m,\rm{loc}}\label{eq:Tij_cont}
\end{align}
where $E^2_{m,\rm{loc}}=E^{i,a}_{m,\rm{loc}}E^{i,a}_{m,\rm{loc}}$ and $E^{i,a}_{m,\rm{loc}}$ and $B^{i,a}_{m,\rm{loc}}$ are local electric and magnetic (cloverleaf) fields. Since the electric field labeled as $E^{i,a}_{m}$ corresponds to the time derivative of the gauge field between $m$ and $m+\hat i$, it is rather centered at the point $m+\hat i/2$. Similarly, a plaquette $U_{ij,m}$ is actually centered at $m+\hat i/2+\hat j/2$. Thus a natural way to construct electric and magnetic fields at a site $m$ is to take nearest neighbor averages to obtain symmetric expressions:
\begin{align}
\label{eq:naiveefieldsymmetric}
    E^{i,a}_{m,\rm{loc}} &{}= \frac{1}{2}\big(E^{i,a}_m + U^\dagger_{i,m-\hat{i}}E^{i,a}_{m-\hat{i}}U_{i,m-\hat{i}}\big)\\
    B^{i,a}_{m,\rm{loc}} &{}= -\frac{\epsilon^{ijk}}{8ga_ja_k} \rm{ReTr}\Big(it^a \big(U_{jk,m} + U_{j-k,m} + U_{-jk,m} + U_{-j-k,m}\big)  \Big)
\end{align}

\subsection{Continuity equation for energy conservation}
\label{sec:E_conserv}

\begin{figure}
    \centering
    \includegraphics[width=0.95\textwidth]{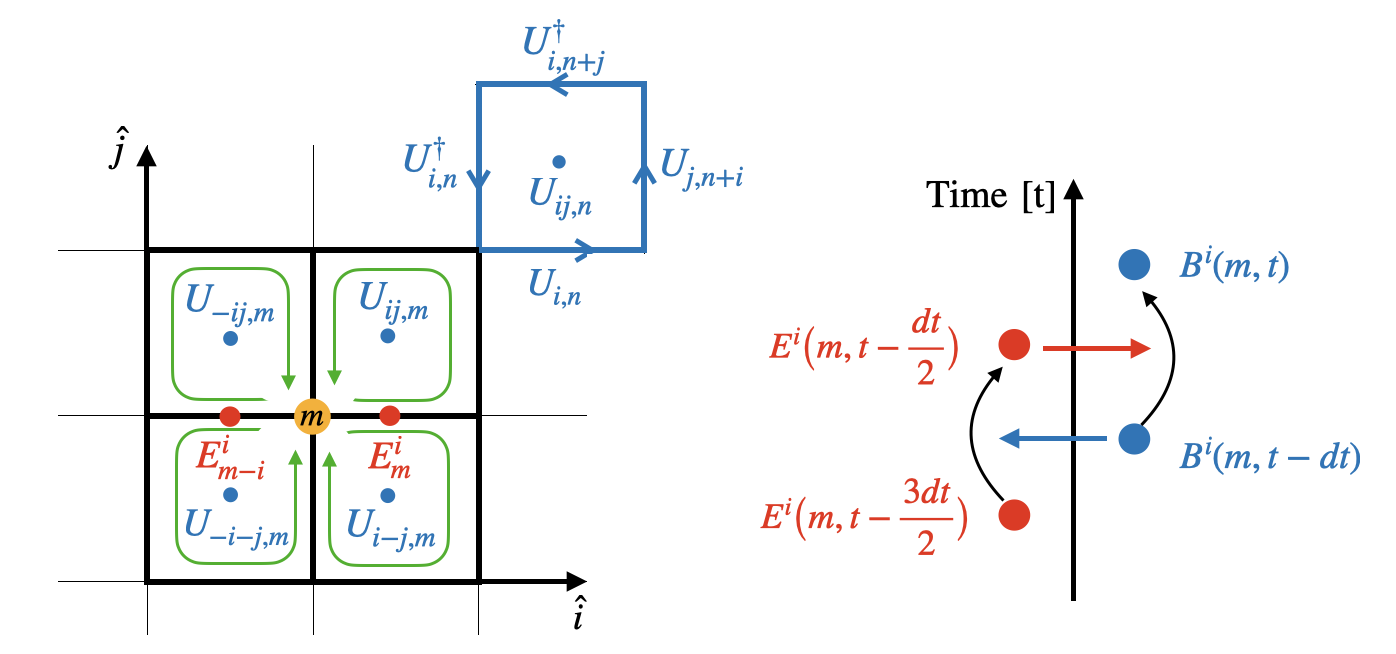}
    \caption{{\em (Left:)} Illustration of discretized $T_{00,m}$ given by Eq.~\eqref{eq:T00_basic}. The black lines with blue central points depict the plaquettes, while red circles represent electric fields. Additionally, the magnetic contribution is illustrated by green arrows indicating the orientation of the plaquettes. The blue and red points indicate the position of the plaquettes and electric fields respectively. The separate highlighted standard plaquette is shown in blue, denoted $U_{ij,n}$ but in fact centered at $n+i/2+j/2$.  
    {\em (Right:)} Presentation of the standard leapfrog algorithm used to update electric and magnetic fields: The electric field at $t-dt/2$ is used to update the magnetic field from $t-dt$ to $t$, and similarly the magnetic field (links) at $t-dt$ to update the electric field from $t-3dt/2$ to $t-dt/2$.
    }
    \label{fig:Lattice_T00}
\end{figure}

Our starting point will be the scalar component of the continuity equation 
\begin{equation}
\label{eq:firstContinuity}
    \partial_0 T_{00} = \partial_i T_{0i},
\end{equation}
which will be used to obtain $T_{0i}$ after $T_{00}$ has been constructed.
As the Hamiltonian density characterizes the energy density of the system, we utilize this insight to identify the $\mu=0$, $\nu=0$ component of the energy-momentum tensor $T_{\mu\nu}$.
It is worth noting that in the Hamiltonian formulation, the electric field labeled $E_{m}^i$ is located at the position $m + \frac{\hat{i}}{2} + \frac{dt}{2}$, taking also the finite timestep $dt$ into account.  The magnetic field strength, expressed by  the plaquette $U_{ij,m}$, is located at $m + \frac{\hat{i}}{2} + \frac{\hat{j}}{2}$. To determine the energy density at lattice site $m$, we compute an average over the ``outgoing'' and ``incoming'' electric fields $E^{j,a}_{m}$ and $E^{j,a}_{m-j}$ and various plaquette orientations. The left panel of Fig.~\ref{fig:Lattice_T00} illustrates the spatial averaging procedure for electric and magnetic contributions.  This averaging procedure allows us to obtain a representative value for the energy density at a specific lattice site
\vspace{0.5em}
\begin{tcolorbox}\vspace{-1em}\begin{align}
    T_{00,m}&{}= \sum_{j,k>0}\frac{1}{g^2a_j^2a_k^2}\Big[N_c-\frac{1}{4}\Big(\rm {ReTr} U_{jk,m} + \rm{ReTr} U_{j-k,m} + \rm {ReTr} U_{-jk,m}+ \rm {ReTr} U_{-j-k,m}\Big)\Big]\notag\\
    &{}+\frac{1}{4}\sum_{j>0}E^{j,a}_mE^{j,a}_m+ E^{j,a}_{m-j}E^{j,a}_{m-j}
    \label{eq:T00_basic}
\end{align}\end{tcolorbox}\noindent

Both the electric and magnetic field components of \eq\eqref{eq:T00_basic} lead to the same total energy as the Hamiltonian \footnote{In Equation \eqref{eq:T00_basic}, we employed a summation over all values of $j$ and $k$ instead of the constrained sum in the Hamiltonian \eqref{eq:CYMhamilton}. The equivalence between the two equations is established by the relation $\sum_{j,k>0}=2\sum_{0<j<k}$.} \eqref{eq:CYMhamilton}, while slightly redistributing the local energy density. This formulation preserves the interpretation of the temporal component of the energy-momentum tensor as energy density.

Taking  a square of the symmetrized electric field from \eq\eqref{eq:naiveefieldsymmetric} as is done in \cite{Ipp:2020nfu, Ipp:2017lho, Schlichting:2020wrv, Matsuda:2020qwx, Avramescu:2023qvv, Matsuda:2023gle} on the other hand, is not equivalent to the actual Hamiltonian \eqref{eq:CYMhamilton}:
\begin{align}
\label{eq:T00E}
\sum_{j,a}\left(E^{j,a}_m + U^\dagger_{j,m-j}E^{j,a}_{m-j}U_{j,m-j}\right)^2 \neq 2\sum_{j,a} \left[E^{j,a}_mE^{j,a}_m+ E^{j,a}_{m-j}E^{j,a}_{m-j} \right].
\end{align}
Furthermore, if one used the square of a symmetrized electric field as on the left-hand side of \eq \eqref{eq:T00E} to calculate the time derivative of $T_{00}$, the equation of motion \eqref{eq:evol_U} would introduce cubic terms in the electric fields into $T_{0i}$, which are not present in the continuum expression. Thus using the right-hand side of \eqref{eq:T00E} is more suitable for the purposes of this paper.

To write the magnetic field part of \eq\eqref{eq:T00_basic} in a form where the equivalence to the Hamiltonian \eqref{eq:CYMhamilton} is more explicit, one needs to write the plaquettes that start from the base point $m$ in \eq\eqref{eq:T00_basic} as plaquettes with a base point at the ``lower left'' corner, parallel transported to the site $m$. This can be done by making use of the identities
\begin{align}
    U_{-kj,m} &{} =(U_{j-k,m})^\dagger=U^\dagger_{k,m-k}U_{jk,m-k}U_{k,m-k} \label{eq:U_{-kj}}\\
    U_{k-j,m} &{} = (U_{-jk,m})^\dagger= U^\dagger_{j,m-j}U_{jk,m-j}U_{j,m-j} \label{eq:U_{k-j}}\\
    U_{-j-k,m} &{}= U^\dagger_{j,m-j}U^\dagger_{k,m-j-k}U_{jk,m-j-k}U_{k,m-j-k}U_{j,m-j} \label{eq:U_{-j-k}}
\end{align}
and the fact that the parallel transporting links cancel in the trace $\rm{Tr}(U^\dag M U) = \rm{Tr}M$. This allows us to rewrite the energy density $T_{00,m}$  as 
\begin{align}
    T_{00,m}&{}= \sum_{j,k>0}\frac{1}{g^2a_j^2a_k^2}\Big[N_c-\frac{1}{4}\Big(\rm {ReTr}\big(U_{jk,m} + U_{jk,m-j} + U_{jk,m-k} + U_{jk,m-j-k}\big)\Big)\Big]\notag\\
    &{}+\frac{1}{4}\sum_{j>0}E^{j,a}_mE^{j,a}_m+ E^{j,a}_{m-j}E^{j,a}_{m-j}\label{eq:T00}.
\end{align}

\subsubsection{Constructing the Poynting vector}

We now want to use the fact that the continuity equation relates the time derivative of the energy density to the spatial derivative of the momentum density to deduce the components $T_{0i}$. To employ this method, we apply the evolution equation for plaquettes \eqref{eq:d0ReTrUjk} and electric fields \eqref{eq:evol_E} to calculate the time derivative of the energy density in Eq.~\eqref{eq:T00}:
\begin{align}
    \partial_0T_{00,m}&{}=\sum_{j,k, j\neq k}\frac{a_ja_k}{2ga_j^2a_k^2}\Big\{\rm{ReTr}\big(i[D_k^F,E^j_m]U_{jk,m} + i[D_k^F,E^j_{m-j}]U_{jk,m-j} + i[D_k^F,E^j_{m-k}]U_{jk,m-k}\notag\\
    &{}+i[D_k^F,E^j_{m-j-k}]U_{jk,m-j-k}\big)+2\rm{ReTr}\big(iE^j_m[D_k^B,U_{jk,m}]+iE^j_{m-j}[D_k^B,U_{jk,m-j}]\big)\Big\}\label{eq:prestep_d0T00}\\
    &{}= \sum_{j,k, j\neq k}\frac{a_j}{2ga_j^2a_k^2}\Big\{\rm{ReTr}\big({iU_{k,m}E^j_{m+k}U^\dagger_{k,m}U_{jk,m}} + {iE^j_mU_{jk,m}}\notag\\
    &{}  + {iU_{k,m-j}E^j_{m-j+k}U^\dagger_{k,m-j}U_{jk,m-j}} + iE^j_{m-j}U_{jk,m-j} - {iU_{k,m-k}E^j_mU^\dagger_{k,m-k}U_{jk,m-k}} \notag\\
    &{} - {iE^j_{m-k}U_{jk,m-k}} - {iU_{k,m-j-k}E^j_{m-j}U^\dagger_{k,m-j-k}U_{jk,m-j-k}} - iE^j_{m-j-k}U_{jk,m-j-k}\big)\Big\}
\end{align}
Indeed, it is possible to write the right-hand side as a total spatial derivative. After some rearrangement, one achieves the following result:
\begin{align} \label{eq:t0kderivation}
    \partial_0T_{00,m}^{(N)}&{}=\sum_{j,k, k\neq j}\frac{1}{2ga_j a_k}\Big\{\rm{ReTr}\big[D_k^B,iU_{k,m}E^j_{m+k}U^\dagger_{k,m}U_{jk,m} + i E^j_mU_{jk,m} \notag\\
    &{}+iU_{k,m-j}E^j_{m-j+k}U^\dagger_{k,m-j}U_{jk,m-j}+iE^j_{m-j}U_{jk,m-j}\big]\Big\}.
\end{align}
Using the definition of the covariant backward (or forward) derivative in \eqs\eqref{eq:defDB}, \eqref{eq:defDF}, it is now easy to see that under the $\rm{ReTr}$ operation, this expression simplifies to an ordinary derivative of a scalar quantity
\begin{equation}
\rm{ReTr} D_{\mu, m}^BX= \frac{1}{a_\mu}\,\rm{ReTr} \big(X_{m}-U^\dagger_{\mu, m-\hat{\mu}}X_{m-\hat{\mu}}~U_{\mu, m-\hat{\mu}}\big)
= 
\frac{1}{a_\mu}\,\rm{ReTr} \big(X_{m}-X_{m-\hat{\mu}}\big)
= \partial_{\mu, m}^B \rm{ReTr} X.
\end{equation}
Thus we have arrived at the first important result of this paper, a discrete energy conservation law
\begin{equation} \label{eq:discrcont}
    \partial_0T_{00,m} = \partial_{k, m}^B    T_{0k,m}.
\end{equation}
We emphasize that this equation is an exact relation even in the discrete case, although its correspondence with the continuum version is only realized in the limit of small lattice spacing. 
From \eq\eqref{eq:t0kderivation} we can read off the momentum density along the $k$-th component as follows:
\vspace{0.5em}
\begin{tcolorbox}\vspace{-1em}\begin{align}
    T_{0k,m}&{}=\sum_{j>0,j\neq k}C'_{jk}\Big\{\rm{ReTr}\big(\underbrace{iE^j_{m+k}U_{-kj,m+k} + i E^j_mU_{jk,m}}_{T_{0k} ^{1}} +\underbrace{iE^j_{m-j+k}U_{-kj,m-j+k}+iE^j_{m-j}U_{jk,m-j}}_{\rm{same~as~} T_{0k} ^{1} \rm{with}~m\rightarrow m-j}\big)\Big\}\label{eq:T_{0k}}
\end{align}\end{tcolorbox}\noindent
where 
\begin{align}
    C'_{jk} = \frac{1}{2 g a_j a_k}
\end{align}
It is important to note that what we label by $T_{0k,m}$ is actually centered at the position $m+\frac{\hat{k}}{2}$,%
\footnote{This can be seen by realizing that the first two terms correspond to the continuum gauge fields in the combination $\left(E^j(m+k+j/2) +E^j(m+j/2)\right)F_{jk}(m+j/2+k/2)$, which is centered around $m+j/2+k/2$, and the second two terms that are shifted as $m\to m-j$ make the full expression centered around $m+k/2$.} 
which may initially seem unconventional. 
However, this arrangement finds justification in the discrete continuity equation \eqref{eq:discrcont}, which has a backward derivative in the $k$-direction. Thus the derivative  $\partial^B_{k, m}  T_{0k}$, is in fact centered around  $m$, the same position where $\partial_0 T_{00}$ is defined. 

We also note that \eq \eqref{eq:T_{0k}} corresponds to a gauge covariant formulation of the Poynting vector in the continuum limit $T^{0k} \to \epsilon^{kij} E^i B^j$. This can be seen by expanding the plaquettes to quadratic order in $a_s$, thus $U_{jk} \approx 1+ i g a_j a_k F_{jk}$, and realizing that the term with the identity vanishes because the matrix $E^j$ is traceless.

\subsubsection{Symmetric time discretization}
\label{sec:symm_time_discr}

Until now, we have treated $t$ as a continuous variable, which is the limit $dt/a_s \ll 1$ of the numerical calculation. In practice, however, one wants to choose a larger timestep for numerical efficiency. In fact, we can make a further improvement to remove some of the finite timestep errors from the conservation law in the following way. To reach this goal
we can further analyze the time dependence of $\partial_0T_{00}$ as described in Eq.~\eqref{eq:T00_basic}. While the Hamiltonian approach we use considers time to be continuous, our numerical simulations do, in fact, have a discrete time step denoted by $dt$. Let us take a closer look at the time dependence of each of the factors and terms in the equation. To simplify the analysis, we will solely focus on the temporal derivative of one of the electric field components in the $T_{00}$ expression 
\begin{align}
\label{eq:timeaverage1}
    \partial_0T_{00,m}^E &{}= \frac{1}{2}\bigg( \frac{E^2_{m}\big(t-\frac{dt}{2}\big) -  E^2_{m}\big(t-\frac{3dt}{2}\big)}{dt}\bigg) \notag\\
    &{}=\bigg(\frac{E^i_m\big(t-\frac{dt}{2}\big) + E^i_m\big(t-\frac{3dt}{2}\big)}{2}\bigg)\bigg(\frac{E^i_m\big(t-\frac{dt}{2}\big) - E^i_m\big(t-\frac{3dt}{2}\big)}{dt}\bigg)\notag\\
    &{}=E^{i,\rm{avg}}_m\big(t-dt\big)\partial_0E^{i}_m\left(t-\frac{dt}{2}\right),
\end{align}
In the leapfrog scheme, the time difference labeled $\partial_0E^{i}_m\big(t-\frac{dt}{2}\big)$ above, corresponding to stepping the electric fields from $t-3dt/2$ to $t-dt/2$, involves the plaquette at the time step $t-dt$, as illustrated in the right panel of \fig \ref{fig:Lattice_T00}. On the other side of the continuity equation, this term corresponds to the one with a spatial derivative of the plaquette, multiplied by the electric field. Equation~\eqref{eq:timeaverage1} tells us that this term in the spatial derivative of $T_{0k}$ should be evaluated with the following timestep assignment in terms of a time-symmetrized electric field:
\begin{align}
\label{eq:timeaverage1b}
    E^j_m[D_k^B,U_{jk,m}] \rightarrow \frac{1}{2}\left(E^j_m \left(t-\frac{dt}{2}\right) + E^j_m \left(t-\frac{3dt}{2}\right) \right)\big[D_k^B,U_{jk,m}(t-dt)\big].
\end{align}
Similarly, the time derivative of the magnetic field part of the energy density corresponds to the term in $\partial_k T_{0k}$ where one takes a spatial derivative of the electric field. A time-symmetric treatment of this term requires a symmetrization of the links in the evaluation of this term in $\partial_k T_{0k}$ as
\begin{align}
\label{eq:timeaverage2}
    [D_k^F,E^j_m]U_{jk,m}\rightarrow \frac{1}{2}\left(\left[D_k^{F}(t),E^j_m\left(t-\frac{dt}{2}\right)\right] + \left[D_k^{F}(t-dt),E^j_m\left(t-\frac{dt}{2}\right)\right]\right)\frac{1}{2}\Big(U_{jk,m}(t) + U_{jk,m} (t-dt) \Big),
\end{align}
where the notation $D_k^{F}(t)$ refers to the link matrices in the covariant derivative being evaluated at the time $t$.  Note that in the temporal gauge, such time-averagings are gauge invariant, since parallel transporters in the time direction are just identity matrices. We refer to the combination of the time averagings \eqref{eq:timeaverage1b} and~\eqref{eq:timeaverage2} as the time-symmetrized discrete formulation.

\subsection{Continuity equation for momentum conservation}
\label{sec:T0i_conserv}

Our goal is to  construct $T^{jk}$ in the same way as we constructed $T^{0i}$ above, i.e., by utilizing the equations of motion and the continuity equation for $i > 0$
\begin{equation}
    \partial_\mu T^{\mu i} = 0
    \label{eq:secondContinuity}
\end{equation}
to derive the $T_{ij}$ components from $\partial_0 T_{0i}$. We first observe that $T_{0k}$ comprises two kinds of terms, with one of them being merely shifted in the $-\hat{j}$ direction. Henceforth, we will focus solely on the first one of the two, which we call $T_{0k}^1$ for our subsequent calculations. The second term can then be restored in the end by shifting the site of $T_{0k}^1$.
Let us first take the time derivative of \eq\eqref{eq:T_{0k}} and split it into terms where the time derivatives act either on the electric fields or the plaquettes.
\begin{align}
    \partial_0 T_{0k,m}^{1} = C'_{jk}\Big\{& \underbrace{E^{j,a}_{m+k}\partial_0\rm{ReTr}(it^a U_{-kj,m+k})+ E^{j,a}_m \partial_0\rm{ReTr} (it^aU_{jk,m})}_{\partial_0T_{0k}^{1E}} \nonumber \\
    + & \underbrace{(\partial_0 E^{j,a}_m) \rm{ReTr} (it^aU_{jk,m}) + (\partial_0E^{j,a}_{m+k})\rm{ReTr}(it^a U_{-kj,m+k})}_{{\partial_0T_{0k}^{1B}}}\Big\}.\label{eq:d0T0K_1}
\end{align}
Since the time derivative of a plaquette \eqref{eq:d0ReTrUjk} involves an electric field, the first term will end up being quadratic in $E$, and reduce in the continuum limit to the electric field part of $T_{ij}$, which we refer to here as $\partial_0T_{0k}^{1E}$. Conversely, the time derivative of the electric field \eqref{eq:evol_E} is a (discrete) spatial derivative of plaquettes, and the second term will end up being quadratic in the magnetic field in the continuum limit, thus denoted by $\partial_0T_{0k}^{1B}$.%
\footnote{These notations should be understood as abbreviated forms of $(\partial_0T_{0k}^{1})^E$ and $(\partial_0T_{0k}^{1})^B$, i.e., the $E$-field and $B$-field parts of the time derivative, rather than time derivatives of $E$-field and $B$-field parts.}

\begin{figure}
    \centering
    \includegraphics[width=0.95\textwidth]{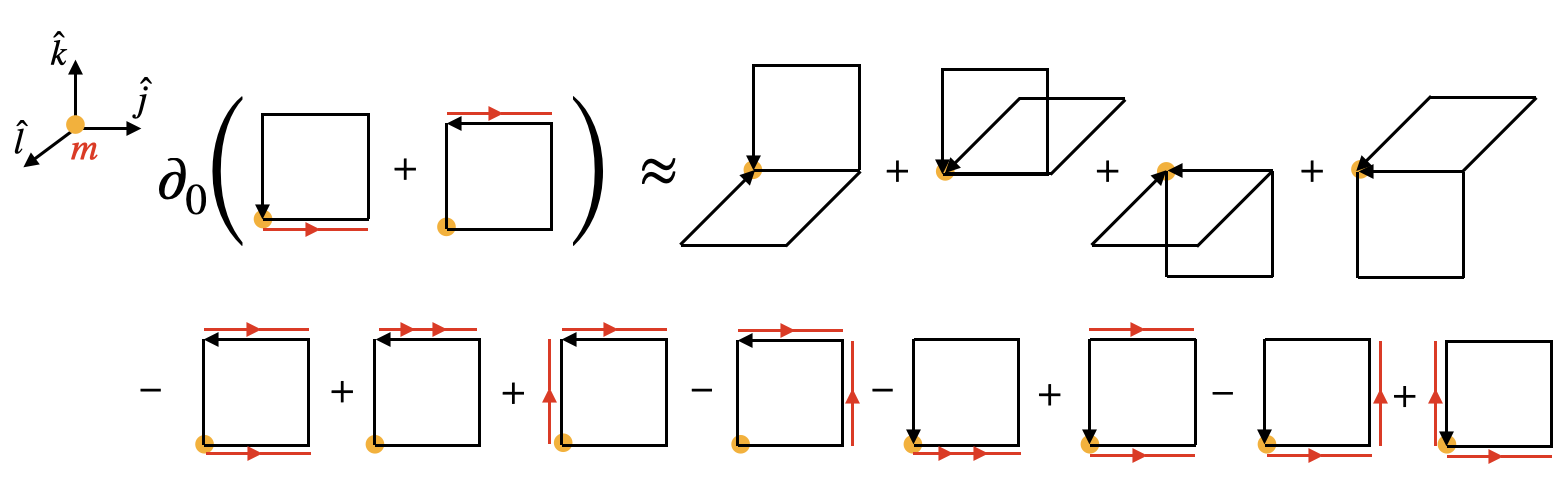}
    \caption{Illustration of the time derivative of $T_{0k}^{1}$ in Eq.~\eqref{eq:T_{0k}}. The filled yellow circle symbolizes the lattice point, while the electric field at a specific point is indicated by red lines.} 
    \label{fig:T0k_challenges}
\end{figure}

\subsubsection{Chromoelectric field contribution}

Let us start with the chromoelectric field contribution~$\partial_0T_{0k}^{1E}$.
We begin by employing the equation of motion for the plaquette to express the electric field part ${\partial_0T_{0k}^{1E}}$ as
\begin{align}
    \partial_0T_{0k,m}^{1E} &{}= \sum_{j\neq k}gC'_{jk}\rm{ReTr}\Big\{a_j\Big(-E^j_{m+k}U^\dagger_{k,m}E^j_mU_{k,m}U_{-kj,m+k} + E^j_{m+k}E^j_{m+k}U_{-kj,m+k}\Big)\notag\\
    &{} + a_k\Big(E^j_{m+k}U^\dagger_{k,m}E^k_mU_{k,m}U_{-kj,m+k} - E^j_{m+k}U^\dagger_{k,m}U_{j,m}E^k_{m+j}U^\dagger_{j,m}U_{k,m}U_{-kj,m+k}
    \Big)\notag\\
    &{} + a_j\Big(-E^j_mE^j_mU_{jk,m} + U_{k,m}E^j_{m+k}U^\dagger_{k,m}E^j_{m}U_{jk,m}\Big) + a_k\Big(-E^j_{m}U_{j,m}E^k_{m+j}U^\dagger_{j,m}U_{jk,m} + E^k_mE^j_mU_{jk,m}\Big)\label{eq:stepA_d0T0k_1E}
    \Big\}.
\end{align}
Here, we have utilized the evolution equation of the plaquette $U_{-kj,m}$, which can be derived in analogy to \eq \eqref{eq:d0ReTrUjk},
\begin{align}
    \partial_0\rm{ReTr}(it^aU_{-kj,m}){}&=ig\rm{ReTr}\Big(it^a a_j\big(U^\dagger_{k,m-k}E^j_{m-k}U_{k,m-k}U_{-kj,m} - U_{-kj,m}E^j_m\big)\notag\\
    &{} - it^a a_k\big(U^\dagger_{k,m-k}E^k_{m-k}U_{k,m-k}U_{-kj,m} - U^\dagger_{k,m-k}U_{j,m-k}E^k_{m+j-k}U^\dagger_{j,m-k}U_{k,m-k}U_{-kj,m}\big) \Big).
\end{align}
Note that in the continuum limit, only the term where the plaquette on the r.h.s is replaced by an identity matrix survives. On the lattice, however, it is not possible to rewrite derivatives of the plaquette in such a way that they would not themselves involve plaquettes. 

By employing this expression in Eq.~\eqref{eq:U_{-kj}}, we can rewrite Eq.~\eqref{eq:stepA_d0T0k_1E} as follows:
\begin{align}
\label{eq:aboveequation}
    \partial_0T_{0k,m}^{1E} &{}= \sum_{j\neq k} gC'_{jk}\rm{ReTr}\Big\{ a_j\Big( E^j_{m+k}E^j_{m+k}U^\dagger_{k,m}U_{jk,m}U_{k,m} - E^j_mE^j_mU_{jk,m}\Big)
    \notag\\&{} +a_k\Big(E^j_{m+k}U^\dagger_{k,m}E^k_mU_{jk,m}U_{k,m} - E^j_{m+k}U^\dagger_{k,m}U_{j,m}E^k_{m+j}U^\dagger_{j,m}U_{jk,m}U_{k,m}\notag\\
    &{} - E^j_mU_{j,m}E^k_{m+j}U^\dagger_{j,m}U_{jk,m} + E^k_mE^j_mU_{jk,m}
    \Big)\Big\}.
\end{align}

In contrast to the previous subsection, where obtaining $T_{0i}$ from $\partial_0T_{00}$ was straightforward, this part is more challenging. In Fig.~\ref{fig:T0k_challenges}, we illustrate the issue by examining the time derivative of $T_{0k}^1$ (see Eq.~\ref{eq:T_{0k}}) at lattice point $m$. The components involving two plaquettes depict the magnetic field component that we will discuss in a later section, while the second line refers to the electric field part $\partial_0T_{0k}^{1E}$. Notably, each term entering the latter includes an extra plaquette, as explicitly stated in Equation~\eqref{eq:aboveequation}. 
This differs from the continuum expression $T_{jk}^{E} = E_jE_k + \frac{1}{2} \delta_{jk}E^2$, where the electric terms are solely quadratic in nature and not entangled with magnetic field components that are encoded in the plaquettes. 

We have not found an exact way of writing $\partial_0T_{0k}^{1E}$ as a total spatial derivative of a quantity that could be identified as a contribution to $T_{ij}$. We, therefore, use the approximation of replacing the plaquettes in $\partial_0T_{0k}^{1E}$ with the identity matrix. This introduces a relative error $\mathcal{O}(a^2)$ and agrees with the original expression in the continuum limit.
After rearranging certain terms and adding the contribution for $\partial_0T_{0k}^{1E}|_{m\rightarrow{m-j}}$ (as in Eq.~\ref{eq:T_{0k}}) the resulting expression takes the following form:
\begin{align}
    \partial_0T_{0k,m}^{E} &{}=  \sum_{j\neq k} gC'_{jk} \rm{ReTr}\Big\{ a_j\Big(E^j_{m+k}E^j_{m+k} - E^j_mE^j_m + E^j_{m+k-j}E^j_{m+k-j} - E^j_{m-j}E^j_{m-j}\Big)-a_ka_j\Big(E^j_m\big[D_j^F,E^k_m\big]
    \notag\\ &{}+ U_{k,m}E^j_{m+k}U^\dagger_{k,m}\big[D_j^F,E^k_m] + E^j_{m-j}\big[D_j^F,E^k_{m-j}\big] +  U_{k,m-j}E^j_{m+k-j}U^\dagger_{k,m-j}\big[D_j^F,E^k_{m-j}] \Big)
    \Big\} \label{eq:d0T_0k_E_step}.
\end{align}
It is important to highlight that when $j$ is equal to $k$, different terms in the first and second lines of $\partial_0T_{0k}^{E}$ cancel each other. This permits us to interchange the summations $\sum_{j\neq k}$ and $\sum_{j,k}$ in the above equation without altering the final outcome. 
Having established this, we will apply the product rule, allowing us to extract the spatial derivative along the $j$ direction. This step is essential to obtain a discretized second (momentum conservation) continuity equation, ultimately enabling us to identify $T_{jk}^E$. The discretized form of the product rule is given as
\begin{align} \label{eq:prodrule}
    \big[D_k^{F}, E^i_mE^j_m\big] = \big[D_k^{F}, E^i_m\big]U_{k,m}E^j_{m+k}U_{k,m} + E^i_m\big[D_k^{F}, E^j_m\big],
\end{align}
where we note the shift from site $m$ to site $m+k$ in the second electric field factor in the first term.
In the case of a parallel transported field, this can be written as 
\begin{equation}
        \big[D_k^{F}, U_{k,m-k}^\dagger E^i_{m-k} U_{k,m-k}E^j_m\big] = [D_k^F, U^\dagger_{m-k}E^i_{m-k}U_{k,m-k}]E^j_m + E^i_m[D_k^F, E^j_m]. 
\end{equation}

We utilize either of these relations to rewrite the terms in Eq.~\eqref{eq:d0T_0k_E_step}, which enables us to employ Gauss' law and eliminate certain contributions from the equation
\begin{align}
   [D_j^B, U^\dagger_{j,m}E^j_mU_{j,m}E^k_{m+j}] &{}= [D_j^F, U^\dagger_{j,m-j}E^j_{m-j}U_{j,m-j}E^k_m] \notag\\
   &{} = E^j_mU_{j,m}E^k_{m+j}U^\dagger_{j,m} - E^j_mE^k_m - U^\dagger_{j,m-j}E^j_{m-j}U_{j,m-j}E^k_m + E^j_mE^k_m\notag\\
   &{}=E^j_{m}[D_j^F,E^k_m] + E^k_m\hspace{-0.2cm}\underbrace{[D_j^B,E^j_m]}_{=~0~ \rm{Gauss~Law}}\label{eq:DiscDer1}
\end{align}
\begin{align}
    [D_j^B,U^\dagger_{j,m}U_{k,m}E^j_{m+k}U^\dagger_{k,m}U_{j,m}E^k_{m+j}] &{} = \big[D_j^F,U^\dagger_{j,m-j}U_{k,m-j}E^j_{m-j+k}U^\dagger_{k,m-j}U_{j,m-j}E^k_m\big]\notag\\
    &{}=U_{k,m}E^j_{m+k}U^\dagger_{k,m}U_{j,m}E^k_{m+j}U^\dagger_{j,m} - U_{k,m}E^j_{m+k}U^\dagger_{k,m}E^k_m\notag\\
    &{} ~~~- U_{k,m-j}E^j_{m+k-j}U^\dagger_{k,m-j}U_{j,m-j}E^k_mU^\dagger_{j,m-j} + U_{k,m}E^j_{m+k}U^\dagger_{k,m}E^k_m\notag\\
    &{}= U_{k,m}E^j_{m+k}U^\dagger_{k,m}\big[D_j^F,E^k_m\big] + \underline{\big[D_j^B,U_{k,m}E^j_{m+k}U^\dagger_{k,m}\big]E^k_m}
    \label{eq:DiscDer2}
\end{align}
\begin{align}
\big[D_j^B,E^j_{m}E^k_{m}\big] &{}=\big[D_j^F,E^j_{m-j}E^k_{m-j}\big]\notag\\
    &{}= E^j_{m-j}[D_j^F,E^k_{m-j}] + E^k_m\hspace{-0.2cm}\underbrace{[D_j^B,E^j_m]}_{= ~0~ \rm{Gauss~Law}}\label{eq:DiscDer3}
\end{align}
\begin{align}
     \big[D_j^B, U_{k,m}E^j_{m+k}U^\dagger_{k,m}E^k_{m}\big]&{}=\big[D_j^F, U_{k,m-j}E^j_{m+k-j}U^\dagger_{k,m-j}E^k_{m-j}\big]\notag\\
    &{} = U_{k,m-j}E^j_{m+k-j}U^\dagger_{k,m-j}\big[D_j^F,E^k_{m-j}\big] + \underline{\big[D_j^B,U_{k,m}E^j_{m+k}U^\dagger_{k,m}\big]E^k_m}\label{eq:DiscDer4}
\end{align}
We can make slight modifications to the underlined terms in Eqns.~\eqref{eq:DiscDer2} and \eqref{eq:DiscDer4} to make use of Gauss' law.
\begin{align}
    \underline{\big[D_j^B,U_{k,m}E^j_{m+k}U^\dagger_{k,m}\big]E^k_m} &{}= U_{k,m}E^j_{m+k}U^\dagger_{k,m}E^k_m - U^\dagger_{j,m-j}U_{k,m-j}E^j_{m+k-j}U^\dagger_{k,m-j}U_{j,m-j}E^k_m\notag\\
    &{}= U_{k,m}E^j_{m+k}U^\dagger_{k,m}E^k_m - U_{-jk,m}U_{k,m}U^j_{m-j+k}E^j_{m-j+k}U_{-kj,m-j+k}U_{j,m-j+k}U^\dagger_{k,m}E^k_m\notag\\
    &{}\simeq U_{k,m}E^j_{m+k}U^\dagger_{k,m}E^k_m - \mathbb{1}U_{k,m}U^j_{m-j+k}E^j_{m-j+k}\mathbb{1}U_{j,m-j+k}U^\dagger_{k,m}E^k_m\notag\\
    &{}=U_{k,m}[D^B_j,E^j_{m+k}]U^\dagger_{k,m}E^k_m \notag\\
    &{}=0 \hspace{1cm}\rm{[Gauss~Law]}
\end{align} 
As we transition from the first to the second equality, we introduce additional gauge links in the second term on the right-hand side, thereby forming plaquettes denoted as $U_{-jk,m}$ and $U_{-kj,m-j+k}$. Subsequently, in the third equality, we approximate the plaquettes with the identity, similarly as discussed earlier, and at the same relative $\mathcal{O}(a^2)$ error, which enables us to apply Gauss' law.
By employing Eqns.~\eqref{eq:DiscDer1} - \eqref{eq:DiscDer4}, we can rephrase Eq.~\eqref{eq:d0T_0k_E_step} as follows:
\begin{align}
    \partial_0T_{0k,m}^{E} &{}=  g \sum_j\sum_lC'_{lj}\delta_{jk}a_la_j \rm{ReTr}\Big\{\big[D_j^B, E^l_{m+j}E^l_{m+j} + E^l_{m+j-l}E^l_{m+j-l}\big]\Big\} - g \sum_jC'_{jk}a_k a_j \rm{ReTr}\Big\{ \big[D_j^B, E^j_{m}E^k_{m} \notag\\
    &{} + U^\dagger_{j,m}U_{k,m}E^j_{m+k }U^\dagger_{k,m}U_{j,m}E^k_{m+j} + U^\dagger_{j,m}E^j_{m}U_{j,m}E^k_{m+j} + U_{k,m}E^j_{m+k}U^\dagger_{k,m}E^k_{m}\big]\Big\}.
\end{align}
This allows us to use the second continuity equation \eqref{eq:secondContinuity}, to identify the electric field component of the stress tensor as follows:
\begin{align}
     T_{jk,m}^{E} &{}= g\bigg\{\sum_l C'_{lj}a_j a_l\delta_{jk} \rm{ReTr}\Big[E^l_{m+j}E^l_{m+j} + E^l_{m+j-l}E^l_{m+j-l}\Big]\notag\\
    &{}-a_ja_kC'_{jk}\rm{ReTr}\Big[U^\dagger_{j,m}U_{k,m}E^j_{m+k }U^\dagger_{k,m}U_{j,m}E^k_{m+j} + E^j_{m}E^k_{m}+U^\dagger_{j,m}E^j_{m}U_{j,m}E^k_{m+j} + U_{k,m}E^j_{m+k}U^\dagger_{k,m}E^k_{m}\Big]\bigg\}.\label{eq:Elec_Tjk}
\end{align}
The contribution $T_{jk}^{E}$ is located at the position\footnote{This is easy to see as follows: the first line of \eq \eqref{eq:Elec_Tjk} is a sum of squares of electric fields leaving from and incoming to point $m+j$ in the $l$-direction, and thus centered around $m+j$, which is equal to $m+j/2+k/2$ for $j=k$. For $j\neq k$ the four terms on the second line are products of the two electric fields associated with each of the four corners of the plaquette $U_{jk,m}$.} $m+\frac{j}{2}+\frac{k}{2}$, precisely as expected from $\partial_{0}T_{0k}$ calculations.

\subsubsection{Chromomagnetic field contribution}
\label{se:magtij}

Having found an expression for the electric contribution in \eqref{eq:Elec_Tjk}, our next objective is to derive an expression for the magnetic field component of $T_{jk}$. However, before delving into that calculation, it is beneficial to examine the continuum expression to illustrate the issues we will encounter along the way. For this discussion to be closer to the discretized version, we will formulate it in terms of the field strength tensor instead of the magnetic field. We begin by computing the magnetic field term in the time derivative of $T_{0i} = E^jF_{ij}$ \footnote{For brevity, we denote in this subsection $A^aB^a = 2 \rm{Tr} A B$ as simply $AB$ for two color matrices $A$ and $B$.}
\begin{align} \label{eq:d0ToiBcont}
    \partial_0 T_{0i}^{B}&{}=[D_j,F_{jq}]F_{iq}.
\end{align}
Comparing this expression to the spatial derivative of the spacelike part of the energy-momentum tensor
$T_{ij}^B = \frac{\delta_{ij}}{4}F_{pq}F_{pq}-F_{iq}F_{jq}$, it is clear that additional terms need to be added to and subtracted from \eqref{eq:d0ToiBcont} to write it as a spatial derivative. In fact, to see how this happens in the continuum, it is easier to start from $T_{ij}^B = \frac{\delta_{ij}}{4}F_{pq}F_{pq}-F_{iq}F_{jq}$ and differentiate it:
\begin{align}
     \partial_j T_{ij}^{B} =  [D_j,T_{ij}^{B}]&{}=-\frac{\delta_{ij}}{2}[D_j,F_{pq}]F_{pq}+[D_j,F_{iq}]F_{jq}+F_{iq}[D_j,F_{jq}]\notag\\
    &{}=-\delta_{ij}[D_j,-\frac{1}{2}\epsilon^{pqr}F_{pq}]B^r+\delta_{ij}[D_j,B^r]B^r-[D_j,B^j]B^i+F_{iq}[D_j,F_{jq}]\notag\\
    &{}=[D_j,F_{jq}]F_{iq}
    \label{eq:djTijB_cont}
\end{align}
In obtaining the last equality, we have utilized the relation $B^r = -\frac{1}{2}\epsilon^{rpq}F_{pq}$ and the Bianchi identity $[D_j, B^j] = 0$. The essential part of this manipulation was the cancellation of the first two terms on the right-hand side of the first line of \eq \eqref{eq:djTijB_cont}:
\begin{align}
    \frac{\delta_{ij}}{2}[D_j,F_{pq}]F_{pq} = [D_j,F_{iq}]F_{jq}\label{eq:ToProove}
\end{align}
Appendix \ref{app:Bianchi} presents an alternative derivation of this same relation.
An important part of the derivation was the Bianchi identity. A version of the Bianchi identity exists on the lattice~\cite{Kiskis:1982ty}. It involves the product of six plaquettes, which, in the continuum, reduces to the continuum Bianchi identity. However, the identity we would need here would involve discretized derivatives of plaquettes, i.e., their nearest-neighbor differences. We have not been able to formulate a suitable exact discrete identity that could play the role of the Bianchi identity in the derivation of the energy-momentum tensor. Thus again, as in the case of the electric field component, we have to resort to an approximation that reduces to \eqref{eq:ToProove} in the continuum limit,
\begin{align}                   
\frac{\delta_{ij}}{2}\rm{ReTr}\Big(it^a\big[D_j^B,U_{pq,m}\big]\Big)\rm{ReTr}\Big(it^aU_{pq,m}\Big) \approx \rm{ReTr}\Big(it^a\big[D_j^B,U_{iq,m}\big]\Big)\rm{ReTr}\Big(it^aU_{jq,m}\Big) .
\label{eq:Disc_Bianchi}
\end{align}
Within this approximation, we are set to calculate the magnetic field contribution of $T_{jk}$ by employing the dynamical equations for the field on the $\partial_0 T_{0k}^{1B}$ component of Eq.~\eqref{eq:d0T0K_1}. These contributions are visualized in the first line of Fig.~\ref{fig:T0k_challenges}, which shows the terms that appear after plugging in the temporal derivatives of the electric fields 
\begin{align}
    \partial_0 T_{0k,m}^{1B} &{} = \sum_{j\neq k} C'_{jk}\Big\{\partial_0E^{j,a}_{m+k}\rm{ReTr}(it^a U_{-kj,m+k}) + \partial_0 E^{j,a}_m \rm{ReTr} (it^aU_{jk,m}) \Big\}\notag\\
    &{}= \sum_{\substack{j\neq k\\j\neq l}}C'_{jk}D_{jl}\Big\{\rm{ReTr}\big(it^a (U_{jl,m+k} + U_{j-l,m+k})\big)\rm{ReTr}\big(it^aU_{-kj,m+k}\big) + \rm{ReTr}\big(it^a(U_{jl,m}+U_{j-l,m})\big)\rm{ReTr}\big(it^aU_{jk,m}\big) 
    \Big\}\notag\\
    &{} =\sum_{l,j}C'_{lk}D_{lj}a_j \Big\{\underbrace{\rm{ReTr}\big(it^a \big[D_j^B,U_{lj,m+k}\big] \big) \rm{ReTr}\big(it^aU_{-kl,m+k}\big)}_{(A)} + \underbrace{\rm{ReTr}\big(it^a\big[D_j^B,U_{lj,m}\big]\big) \rm{ReTr}\big(it^aU_{lk,m}\big)}_{(B)}\Big\},\label{eq:step1_d0T0k1B}
\end{align}
where 
\begin{align}
    D_{jl}=\frac{2a_j}{ga_j^2a_l^2}.
\end{align}
Transitioning from the second equality to the third, we took into account $\rm{ReTr}(it^a\mathds{1}) = 0$ under the conditions $j =k$ and $j=l$, and subsequently exchanged the indices $j$ and $l$.

By employing the Fierz identity $t^a_{ij}t^a_{kl} = \frac{1}{2}\big(\delta_{il}\delta_{jk}- \frac{1}{N_c}\delta_{ij}\delta_{kl}\big)$ for the generators in the fundamental representation, the multiplication of two arbitrary plaquettes can be streamlined. 
Let us at this point simplify our notations and leave out contributions arising from the $N_c$-suppressed term in the Fierz identity. For SU(2) this term is in fact zero because the trace of a unitary matrix is purely real. For SU(3) the trace of a plaquette can have an imaginary part, but it is suppressed by powers of the lattice spacing. Since we are in any case not obtaining an expression that is exact at all orders in the lattice spacing, we will not write these trace terms here. With these approximations, we obtain 
\begin{align}
    \rm{ReTr}(it^aU_{\mu\nu,m})\rm{ReTr}(it^aU_{\alpha\beta,m})&{} \approx \frac{i^2}{4}\rm{ReTr}\big(U_{\mu\nu,m}U_{\alpha\beta,m} -U_{\mu\nu,m}U_{\beta\alpha,m}\big),\label{eq:FierzIdentity}
\end{align}
where we have utilized the fact that the anti-Hermitian part of the $SU(2)$ matrix is traceless to remove the $1/N_c$ term from the equation.
Now, with the approximation \eqref{eq:FierzIdentity}, part $(A)$ of Eq.~\eqref{eq:step1_d0T0k1B} can be rewritten as 
\begin{align}
    (A) &{} =\frac{i^2}{4}\rm{ReTr}\big(U_{-kl,m+k} \big[D_j^B, U_{lj,m+k}-U_{jl,m+k}\big]\big)\notag\\
    &{} = \frac{i^2}{4}\Big\{ \rm{ReTr}\big(U_{-kl,m+k}\big[D_j^B,U_{lj,m+k}\big] + \big[D_j^B, U_{j,m+k}U_{-kl,m+k+j}U^\dagger_{j,m+k}\big]U_{lj,m+k}
    \notag\\
    &{}-\big[D_j^B, U_{j,m+k}U_{-kl,m+k+j}U^\dagger_{j,m+k}\big]U_{lj,m+k} - U_{-kl,m+k}\big[D_j^B,U_{jl,m+k}\big] \notag\\
    &{}- \big[D_j^B, U_{j,m+k}U_{-kl,m+k+j}U^\dagger_{j,m+k}\big]U_{jl,m+k} + \big[D_j^B, U_{j,m+k}U_{-kl,m+k+j}U^\dagger_{j,m+k}\big]U_{jl,m+k}\Big\}.
\end{align}
Here, we employed insights from our prior discussion on the continuum case to add and subtract terms strategically. By utilizing the discretized product rule, certain terms can be combined
\begin{align}
    (A)&{} =\frac{i^2}{4}\Big\{\rm{ReTr}\Big(\big[D_j^B, U_{j,m+k}U_{-kl,m+k+j}U^\dagger_{j,m+k}U_{lj,m+k}\big]
    -\big[D_j^B, U_{j,m+k}U_{-kl,m+j+k}U^\dagger_{j,m+k}U_{jl,m+k}\big]\notag\\
    &{}-\big[D_j^B,U_{j,m+k}U_{-kl,m+k+j}U^\dagger_{j,m+k}\big]U_{lj,m+k}+\big[D_j^b,U_{j,m+k}U_{-kl,m+k+j}U^\dagger_{j,m+k}\big]U_{jl,m+k} 
    \Big)\Big\}\notag\\
    &{}=\frac{i^2}{4}\rm{ReTr}\Big(\big[D_j^B,U_{j,m+k}U_{-kl,m+j+k}U^\dagger_{j,m+k}\big(U_{lj,m+k}-U_{jl,m+k}\big)\big]\notag\\
    &{}-\big[D_j^B, U_{j,m+k}U^\dagger_{k,m+j}U_{lk,m+j}U_{k,m+j}U^\dagger_{j,m+k}\big]\big(U_{lj,m+k}-U_{jl,m+k}\big)\Big),
\end{align}
while the remaining terms can be substituted with the approximate lattice Bianchi Identity in Eq.~\eqref{eq:Disc_Bianchi} to give the final expression as
\begin{align}
    (A) &{} \approx \frac{i^2}{4}\rm{ReTr}\Big(\big[D_j^B, U_{j,m+k}U_{-kl,m+k+j}U^\dagger_{j,m+k} \big(U_{lj,m+k}-U_{jl,m+k}\big)\big] -\frac{\delta_{jk}}{4}\big[D_j^B, U_{pl,m+k}\big(U_{pl,m+k}-U_{lp,m+k}\big)\big]\Big).\label{eq:part1_d0T0k}
\end{align}
With this, we employ a similar approach for the term labeled as $(B)$ in Eq.~\eqref{eq:step1_d0T0k1B}
\begin{align}
    (B) &{}= \frac{i^2}{4}\rm{ReTr}\Big(U_{lk,m}\big[D_j^B, U_{lj,m}-U_{jl,m}\big]\Big)\notag\\
    &{}=\frac{i^2}{4}\rm{ReTr}\Big(U_{lk,m}\big[D_j^B, U_{lj, m}\big] + U^\dagger_{j,m-j}U_{lj,m-j}U_{j,m-j}\big[D_j^B, U_{lk,m}\big] - U^\dagger_{j,m-j}U_{lj,m-j}U_{j,m-j}\big[D_j^B, U_{lk,m}\big] \notag\\
    &{}- U_{lk,m}\big[D_j^B, U_{jl,m}\big] - U^\dagger_{j,m-j}U_{jl,m-j}U_{j,m-j}\big[D_j^B,U_{lk,m}\big] + U^\dagger_{j,m-j}U_{jl,m-j}U_{j,m-j}\big[D_j^B,U_{lk,m}\big]\Big)\notag\\
    &{}=\frac{i^2}{4}\rm{ReTr}\Big(\big[D_j^B, U_{lk,m}U_{lj,m} - U_{lk,m}U_{jl,m}\big] + \big[D_j^B,U_{lk,m}\big] \big(U_{l-j,m} - U_{-jl,m}\big)\Big)\notag\\
    &{}\approx \frac{i^2}{4}\rm{ReTr}\Big(\big[D_j^B, U_{lk,m}\big(U_{lj,m} - U_{jl,m}\big)\big] -\frac{\delta_{jk}}{4} \big[D_j^B, U_{pl,m}\big(U_{pl,m}-U_{lp,m}\big)\big]\Big) \Big).\label{eq:part2_d0T0k}
\end{align}
By combining the aforementioned equations \eqref{eq:part1_d0T0k} and \eqref{eq:part2_d0T0k}, along with their shifted counterparts $\partial_0 T_{0k}^{1B}|_{m\rightarrow{m-j}}$ in Eq.~\eqref{eq:T_{0k}}, we derive the approximate magnetic contribution of the stress tensor as
\begin{align}
    T_{jk,m,CB}^{B} &{} \approx\sum_{l}C'_{lk}D_{lj}\frac{a_j}{4}\rm{ReTr}\bigg\{-U_{j,m+k}U^\dagger_{k,m+j}U_{lk,m+j}U_{k,m+j}U_{j,m+k}^\dagger\big(U_{lj,m+k}- U_{jl,m+k}\big) - U_{lk,m}\big(U_{lj,m}-U_{jl,m}\big)\notag\\
    &{}+\frac{\delta_{jk}}{4}\Big(U_{pl,m+k}\big(U_{pl,m+k} - U_{lp,m+k}\big) + U_{pl,m}\big(U_{pl,m}-U_{lp,m}\big)\Big)\notag\\
    &{}-U_{j,m+k-l}U^\dagger_{k,m+j-l}U_{lk,m+j-l}U_{k,m+j-l}\big(U_{lj,m+k-l}- U_{jl,m+k-l}\big) - U_{lk,m-l}\big(U_{lj,m-l}-U_{jl,m-l}\big)\notag\\
    &{}+\frac{\delta_{jk}}{4}\Big(U_{pl,m+k-l}\big(U_{pl,m+k-l} - U_{lp,m+k-l}\big) + U_{pl,m-l}\big(U_{pl,m-l}-U_{lp,m-l}\big)\Big)
    \bigg\}.\label{eq:Mag_Tjk}
\end{align}
Here, the notation $T_{jk, m, CB}^B$ is employed to represent the magnetic-field component of the stress tensor derived from the conjectured Bianchi (CB) identity. Note that in order to write the $\delta_{jk}$-part of $T_{jk}^B$ explicitly in this form, we have used the continuum form of the product rule, rather than the exact one \eqref{eq:prodrule}, and the approximate Bianchi identities \eqref{eq:Disc_Bianchi}. 

\begin{figure}
    \centering
    \includegraphics[width=0.4\textwidth]{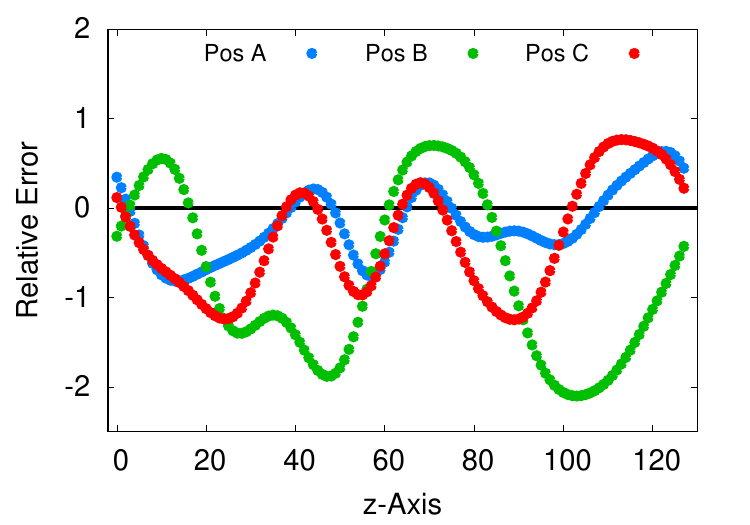}
    \caption{Relative error \eqref{eq:BianchiRelativeError} in the conjectured Bianchi identity at different lattice sites in the transverse plane (A, B and C) as a function of the longitudinal direction}
    \label{fig:Rel_error_Bianchi}
\end{figure}

We have checked the quality of the conjectured approximate Bianchi identity \eqref{eq:Disc_Bianchi} approximation numerically, using the setup that will be discussed in more detail in Sec.~\ref{se:numerics}. The result is demonstrated in
 Fig.~\ref{fig:Rel_error_Bianchi}. Here  we show the relative error in the lattice counterpart of the conjectured Bianchi identity, calculated as 
\begin{align}
\label{eq:BianchiRelativeError}
\frac{\rm{ReTr}\big(it^a\big[D_j^B,U_{lk,m}\big]\big)\rm{ReTr}\big(it^aU_{lj,m}\big) - \frac{\delta_{jk}}{2}\rm{ReTr}\big(it^a\big[D_j^B,U_{pl,m}\big]\Big)\rm{ReTr}\big(it^aU_{pl,m}\big)}{\frac{1}{V}\sum_{x}\rm{ReTr}\big(it^a\big[D_j^B,U_{lk,m}\big]\big)\rm{ReTr}\big(it^aU_{lj,m}\big)}
\end{align}
at distinct lattice sites labeled as A, B and C within the transverse plane. It is evident that the relative error is large and exhibits local variations, thereby suggesting the potential for devising an improved representation of the lattice Bianchi identity in future studies.

For now, we need an alternative, more accurate expression. Since our conjectured Bianchi identity is not satisfied well for the field configurations that we would like to apply it to, we will use another approach that circumvents the main issues.
Here, one first notices that in our derivation leading to $T_{jk}^B$, the Bianchi identity has been used only to manipulate the total energy contribution that is proportional to $\delta_{jk}$. In the continuum, the magnetic field part of the $\delta_{jk}$-term is just the same magnetic field squared that appears in the energy density. Thus, we will follow another approach that does not rely on \eq\eqref{eq:Disc_Bianchi}, which is to just take the magnetic field part of the total energy density $T_{00}^B$ and use it to get the magnetic $\delta_{jk}$ part of $T_{jk}$. This leads to 
\begin{align}
    T_{jk,m}^{B} &{}= \sum_{l}C'_{lk}D_{lj}\frac{a_j}{4}\rm{ReTr}\bigg\{-U_{j,m+k}U^\dagger_{k,m+j}U_{lk,m+j}U_{k,m+j}U_{j,m+k}^\dagger\big(U_{lj,m+k}- U_{jl,m+k}\big) - U_{lk,m}\big(U_{lj,m}-U_{jl,m}\big)\notag\\
    &{} - U_{j,m+k-l}U^\dagger_{k,m+j-l}U_{lk,m+j-l}U_{k,m+j-l}\big(U_{lj,m+k-l}- U_{jl,m+k-l}\big) - U_{lk,m-l}\big(U_{lj,m-l}-U_{jl,m-l}\big)\bigg\} \nonumber\\
    &{} - \delta_{jk}T_{00, m+j}^B. \label{eq:Mag_Tjk_wed}
\end{align}
This is achieved by subtracting $T_{00,m+j}^B$ based on the first term of the continuum expression ${T_{ij} = -1/4 \delta_{ij}F_{mn}F_{mn} + F_{in}F_{jn}}$. Additionally, this contribution from the energy density needs to be taken at the point $m+j$, as $T_{jk}$ for $j=k$ needs to be situated at the point $m+j$ in order to yield $\partial_0 T_{0k}$ centered at $m+k/2$  as a backward derivative in the $j=k$ direction. Equation \eqref{eq:Mag_Tjk_wed} thus does not use the approximate Bianchi identity and turns out to introduce only a small relative error to the continuity equation, as we will show below.

\subsubsection{Final formula}

With this, we write our final expression for $T_{jk}$ by unifying the electric  and magnetic components \eqref{eq:Elec_Tjk} and \eqref{eq:Mag_Tjk_wed}
\vspace{0.5em}
\begin{tcolorbox}\vspace{-1em}\begin{align}\label{eq:final_Tjk} 
    T_{jk,m} &{}= \bigg\{\sum_l gC'_{lj}a_j a_l\delta_{jk} \rm{ReTr}\Big[E^l_{m+j}E^l_{m+j} + E^l_{m+j-l}E^l_{m+j-l}\Big]\notag\\
    &{}-ga_ja_kC'_{jk}\rm{ReTr}\Big[U^\dagger_{j,m}U_{k,m}E^j_{m+k }U^\dagger_{k,m}U_{j,m}E^k_{m+j} + E^j_{m}E^k_{m}+U^\dagger_{j,m}E^j_{m}U_{j,m}E^k_{m+j} + U_{k,m}E^j_{m+k}U^\dagger_{k,m}E^k_{m}\Big]\bigg\}\notag\\
    &{} +\sum_{l}C'_{lk}D_{lj}\frac{a_j}{4}\rm{ReTr}\bigg\{-U_{-kl,m+j+k}\big(U_{-jl,m+j+k}- U_{l-j,m+j+k}\big) - U_{lk,m}\big(U_{lj,m}-U_{jl,m}\big)\notag\\
    &{}- U_{-kl,m+j+k-l}\big(U_{-jl,m+j+k-l}- U_{l-j,m+j+k-l}\big) - U_{lk,m-l}\big(U_{lj,m-l}-U_{jl,m-l}\big)\bigg\} - \delta_{jk}T_{00, m+j}^B .
\end{align}\end{tcolorbox}\noindent

\begin{figure}
    \centering
    \includegraphics[width=0.45\textwidth]{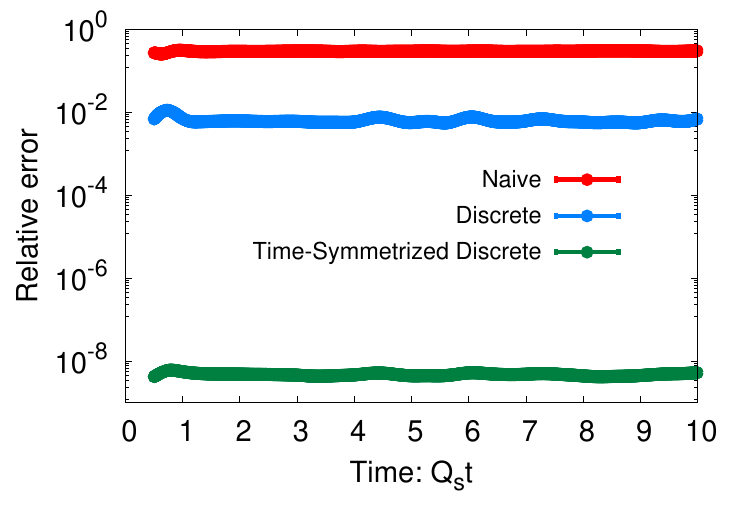}
    \includegraphics[width=0.45\textwidth]{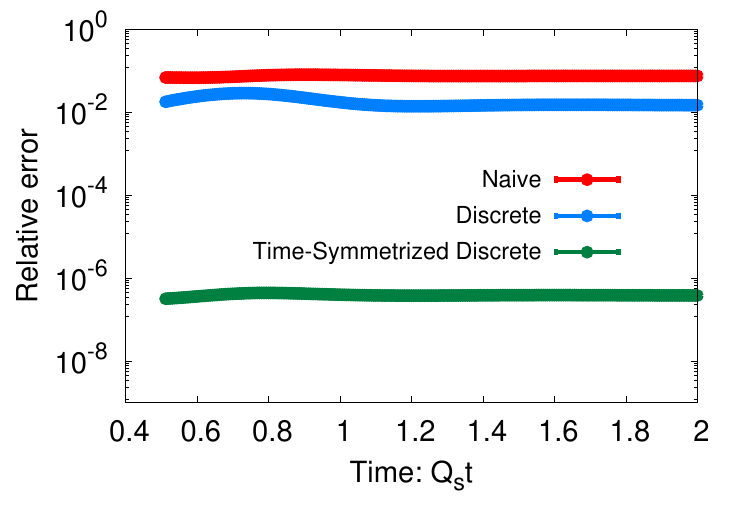 }
    \includegraphics[width=0.45\textwidth]{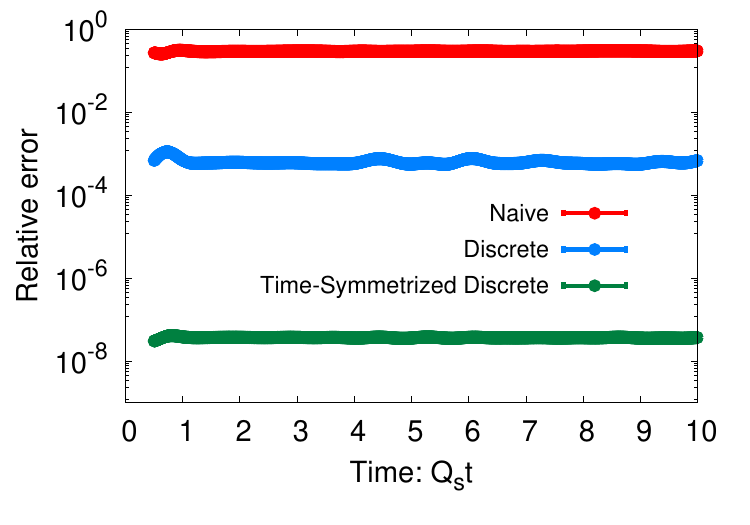}
    \includegraphics[width=0.45\textwidth]
    {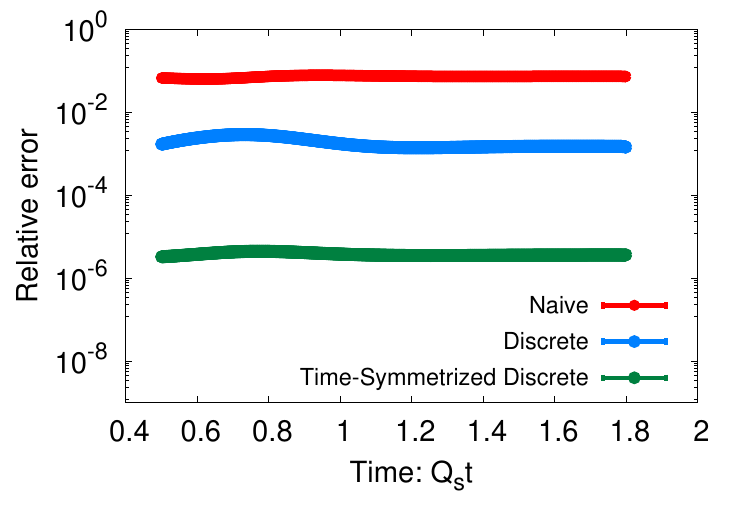}
    \caption{The relative error \eqref{eq:EnergyRelError} in the energy conservation equation is shown as a function of time on the left for $N_s=32$ and on the right for $N_s=128$. The lines correspond to different formulations of the Poynting vector using the `naive' continuum \eqref{eq:T0i_cont}, `discrete' \eqref{eq:T_{0k}} and time-symmetrized (obtained using the procedure described by \eqref{eq:timeaverage1} and \eqref{eq:timeaverage2}) expressions. The upper panels represent $dt/a_s$ values of 0.01 and 0.1 for $N_s=32$ and $a_s = 0.25$ and for $N_s=128$ and $Q_s a_s = 0.0625$, respectively, while the lower panels employ 0.001 and 0.01, respectively.
    }
    \label{fig:cont_eq_1}
\end{figure}

\section{Numerical Results}\label{se:numerics}

\subsection{Initial conditions and numerical setup}

To quantitatively assess our expressions for the energy-momentum tensor, we test the conservation laws in a classical Yang-Mills simulation. In our numerical simulations, we solve the evolution equations \eqref{eq:evol_U} and \eqref{eq:evol_E}  for the lattice fields using the standard leapfrog algorithm, and then calculate components of the energy-momentum tensor. 
At the beginning, we take the fields from  initial conditions of the form 
\begin{align}
    \label{eq:init_AA}
    \langle A_i^a(\mathbf p,t{=}0) A_i^{*b}(\mathbf p,t{=}0)  \rangle &= 2 a_s^3 N_s^3 \delta_{ab}\,\frac{0.2}{g^2}\,\frac{Q_s}{p^2}\,e^{-p^2/2Q_s^2} \\
    \langle A_i^a(t=0, p) \rangle &= 0 \\
    E_i^a(t=0) &= 0,
\end{align}
where $g$ is the coupling constant, $Q_s$ a dimensionful initial momentum scale, and $a_s^3 N_s^3$ the lattice volume. The expectation value $\langle \cdot \rangle$ implies an average over classical configurations of the lattice fields.
In \eq \eqref{eq:init_AA} we only initialize transverse modes, $p^i A_i(\mathbf p, t{=}0) = 0$.
With these initial conditions, all of the energy of the system resides initially in the chromomagnetic fields, and electric fields are generated over the time evolution of the system. These initial conditions are the magnetic field part of the ones used, e.g., in \cite{Boguslavski:2018beu, Boguslavski:2020tqz}. Choosing the initial condition to have a zero electric field allows Gauss' law to be exactly satisfied at the initial condition, while the leapfrog algorithm for the time evolution preserves it to machine precision.%
\footnote{The exact conservation of Gauss' law is crucial for conserved energy-momentum tensor since our derivation of $T_{jk}$ in \eq \eqref{eq:final_Tjk} heavily relies on it. For the purposes of this paper, we have chosen to use an initial condition that satisfies Gauss' law exactly, up to accumulated errors from floating point calculations. Future studies performed using more realistic conditions will, however, have to pay special attention to the accuracy of Gauss' law.}
After the characteristic timescale $t\sim 1/Q_s$ the energy becomes roughly evenly distributed between the electric and magnetic fields. 

Within this setup, we will present our results for lattices with a constant physical volume $N_s a$, where we change the lattice spacing $a_s = \{0.25, 0.125, 0.0625, 0.03125\}$ and the lattice size $N_s = \{32, 64, 128, 256\}$ simultaneously.
We will also vary the time step $dt$. Due to computational complexity, we will conduct all our simulations using the $SU(2)$ gauge group instead of the physical $SU(3)$ gauge group of QCD.

\begin{figure}
    \centering
    \includegraphics[width=0.45\textwidth]{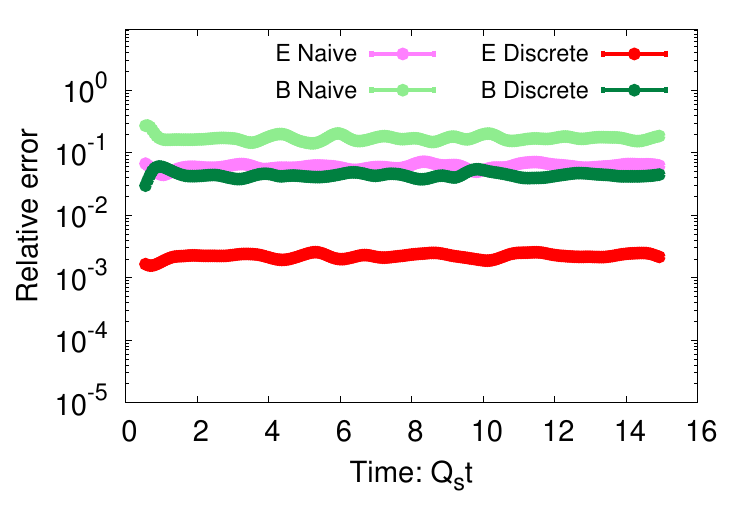}
    \includegraphics[width=0.45\textwidth]{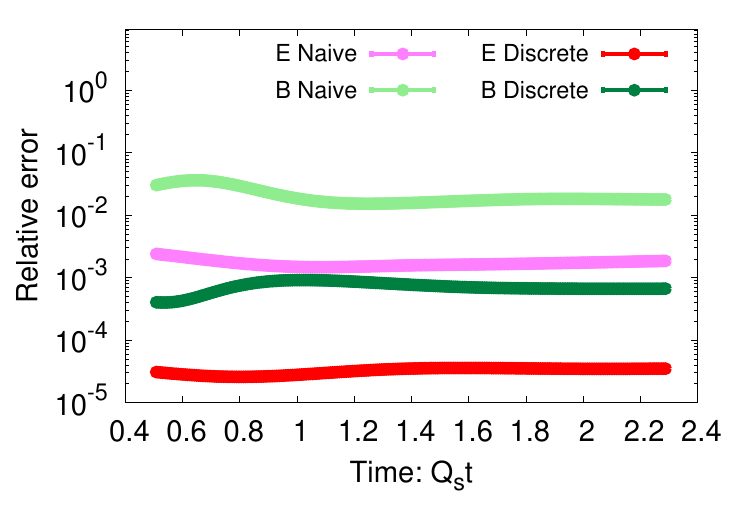}
    \caption{Relative error \eqref{eq:relerror} in the momentum conservation equation as a function of time for $N_s=32$  on the left and $N_s=256$ on the right. Both are shown for the electric and magnetic field contributions in the continuum \eqref{eq:Tij_cont} and discretized \eqref{eq:final_Tjk} formulations.}
    \label{fig:cont_eq_2}
\end{figure}

\subsection{Results}

Based on the above expressions \eqref{eq:T00_basic}, \eqref{eq:T_{0k}} and \eqref{eq:final_Tjk} of the energy-momentum tensor, we will now study the violation of the continuity equations $\partial_\mu T^{\mu\nu} = 0$. We start our analysis with the equation for energy conservation, where we quantify the deviation in the form of a relative error
\begin{align}
\label{eq:EnergyRelError}
    \sqrt{\frac{\sum\limits_{m} \big(\partial_0 T_{00,m} -\partial_{i}T_{0i,m}\big)^2}{\sum\limits_{m} \big(\partial_0 T_{00,m}\big)^2}}\,.
\end{align}
Here, the time derivative is calculated explicitly as a time difference between two discrete timesteps. 

Figure~\ref{fig:cont_eq_1} illustrates the evolution of the relative error in three scenarios: one employing the straightforward ``naive'' discretization method outlined in \eqs \eqref{eq:T00_cont} and~\eqref{eq:T0i_cont}, and the others employing the discretized approach specified in \eq \eqref{eq:T00_basic} along with the Poynting vector formulated either by \eq \eqref{eq:T_{0k}} or the time-symmetrized discretization outlined in \se \ref{sec:symm_time_discr}. The results are presented for a spatially three-dimensional lattice with $N_s=32$ in the left and $N_s=128$ in the right panels. The top panel corresponds to $dt/a_s$ values of 0.01 and 0.1 for $N_s=32$ and $128$, respectively. The lower panels illustrate the relative error for one-tenth of these values for $dt/a_s$. First of all, it is evident that the violation remains relatively consistent over time, even after the energy becomes distributed over both electric and magnetic fields. Secondly, the discrete approach with time symmetrization exhibits exceptional performance for different lattice spacings and timesteps. Thirdly, for smaller lattices, the discrete method outperforms the naive description. However, as we approach the continuum limit by increasing $N_s$ and decreasing the lattice spacing accordingly, the naive description improves and approaches the discrete one as expected. 

When examining the upper panel of \fig \ref{fig:cont_eq_1} in comparison to the lower one with a smaller $dt/a_s$ ratio by a factor of ten, one observes that the relative error of the discrete approach reduces roughly by the same factor. This is again consistent with the expectation that the discrete formulation in this regime is dominated by timestep effects. However, we note that the time-symmetric discrete method is still several orders of magnitude more precise even with a rather small timestep.
The error in the time-symmetrized formulation does not decrease anymore when the timestep is decreased from the top row to the bottom, which we take as an indication that it has already reached a limit where it is dominated by machine precision effects. 

In Fig.~\ref{fig:cont_eq_2}, we examine the relative error of the second continuity equation \eqref{eq:secondContinuity}, quantified by
\begin{align}
\label{eq:relerror}
    \sqrt{\frac{\sum\limits_{m} \big(\partial_0 T_{0i,m}^{E/B} -\partial_{j}T_{ij,m}^{E/B}\big)^2}{\sum\limits_{m} \big(\partial_0 T_{0i,m}^{E/B}\big)^2}}
\end{align}
as a function of time for $i=1$. The violation is measured separately for the electric and magnetic field components for $N_s=32$ on the left and $N_s=256$ on the right-hand side as before. The naive approach is based on \eqs \eqref{eq:T0i_cont} and \eqref{eq:Tij_cont}, while the discrete approach utilizes \eqs \eqref{eq:T_{0k}} and \eqref{eq:final_Tjk}.
In the figure, one observes that the discrete formulation surpasses the naive approach. This effect is particularly large for the electric field part but is also present for the magnetic field contribution. 
Indeed, the violation of the electric contribution in the discrete case is roughly an order of magnitude smaller than for the magnetic one, which indicates that our approximations for the electric sector of $T_{ij}$ are more accurate than for the magnetic one.
As we approach the continuum limit by considering the right panel, the violation further diminishes for both the naive and the discretized expressions, while other trends remain consistent.
Thus, while we have not been able to achieve a description with an exact conservation law on a discrete lattice, we have managed to find a formulation where the lattice discretization effects on the conservation law are significantly smaller than for a naive discretization.

\begin{figure}
    \centering
    \includegraphics[width=0.45\textwidth]{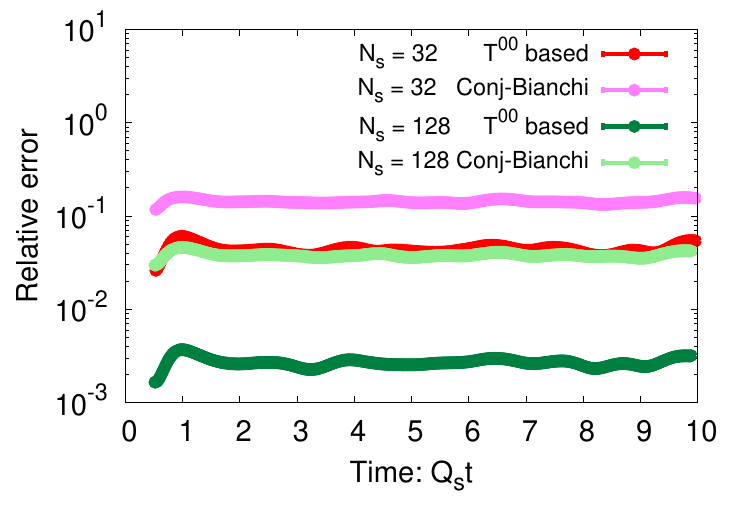}
    \caption{Measuring the violation \eqref{eq:relerror} in the second continuity equation \eqref{eq:secondContinuity} by utilizing $T_{ij}^B$ derived from the energy density $T_{00}$ \eqref{eq:final_Tjk} and the conjectured Bianchi identity \eqref{eq:Mag_Tjk}, considering two distinct lattice sizes: $N_s=32$ and $N_s=128.$ }
    \label{fig:Bianchi_comparison}
\end{figure}
\begin{figure}
    \centering
    \includegraphics[width=0.45\textwidth]{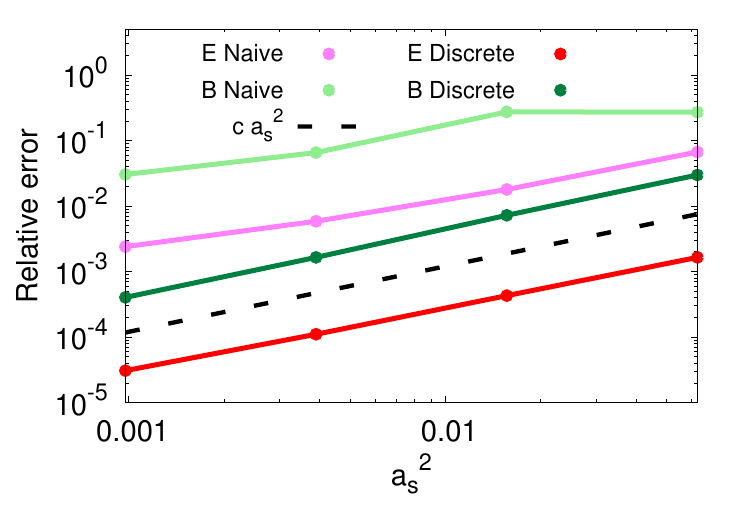}
    \caption{Relative error \eqref{eq:relerror} in the second continuity equation \eqref{eq:secondContinuity} as a function of lattice spacing $a_s^2$ for discretized \eqref{eq:final_Tjk}  and naive (continuum limit) electric and magnetic fields  \eqref{eq:Tij_cont}. Black dashed lines correspond to the $a_s^2$ power-law.  }
    \label{fig:SpacingDep}
\end{figure}

As discussed in Sec.~\ref{se:magtij}, we had two formulations for the magnetic part of $T_{ij}$, one based on a conjectured approximation that reduces to the Bianchi identity in the continuum, the other one using the discrete energy density  $T^{00}$ to reconstruct the part of  $T_{ij}$ that is proportional to $\delta_{ij}$.
Figure~\ref{fig:Bianchi_comparison} offers a comprehensive analysis of the difference between these two approaches for the violation of the momentum conservation equation, as parametrized by the relative error \eqref{eq:relerror}.
We investigate the evolution of the relative error for lattice sizes of $N_s=32$ and $N_s=128$ for those two distinct stress tensor formulations. 
We note that the expression based on $T_{00}$ \eqref{eq:final_Tjk} performs significantly better than the one derived utilizing the lattice analog of the Bianchi identity \eqref{eq:Mag_Tjk}, demonstrating an improvement of at least one order of magnitude. As we approach the continuum limit with $N_s=128$, we observe a decrease in violation for both cases, as anticipated. Notably, the reduction occurs at a more favorable rate for the $T_{00}$-based approach. This justifies our choice of the approach chosen in \eqref{eq:final_Tjk}. In future research, it would be beneficial to develop a more refined expression for the Bianchi identity to ensure that the derivation of a discretized $T_{ij}$ aligns seamlessly with the continuum case. 

The denominator of the relative error \eqref{eq:relerror} can be understood as the space-integrated squared rate of change of energy flux in the $i$-direction. Thus it corresponds to a physical quantity, which takes a nonzero value at the continuum limit. Hence, the relative error should tend to zero in the continuum limit, and the $a_s$ scaling of the ratio is that of the numerator (discretization errors of the denominator represent a subleading correction to this behavior). Figure \ref{fig:SpacingDep} illustrates this quantity for the naive and improved discretizations as functions of the quadratic lattice spacing $a_s^2$. The black dashed line corresponds to a power law $\sim a_s^2$. We observe that in the limit of small lattice spacing, the relative error of the naive expression goes to zero slower than $a_s^2$. In contrast to this, the improved discrete expressions follow the $a_s^2$ power law very closely, thus indicating that they are $\mathcal{O}(a_s^2)$.

The discretized version of the second continuity equation \eqref{eq:secondContinuity} would also benefit from the time-symmetrizing procedure that was performed on $T^{00}$ and $T^{0i}$ above. However, while deriving \eq \eqref{eq:Mag_Tjk_wed}, we have already performed an error $\mathcal{O}(a_s^2)$ when replacing plaquettes with identity operators. Similarly, the approximations in the magnetic sector are at most of the same accuracy. Artifacts arising from the time discretization are typically subleading to the spatial discretization effects since we employ $dt / a_s \ll 1$ to guarantee numerical stability. Hence, we do not consider corrections to time discretization for the momentum conservation equation.

\section{Conclusions}
\label{se:conclu}

The use of numerical simulations to gain insights into nonperturbative aspects of classical and quantum field theories requires discretizing space-time on a lattice and demands a systematic approach to study the energy-momentum tensor for nonabelian gauge fields.
We present an improved expression for the this purpose, given by  Eqs.~\eqref{eq:T00_basic}, \eqref{eq:T_{0k}}, and \eqref{eq:final_Tjk}, that are derived using classical field equations of motion in conjunction with the energy-momentum conservation law for $T^{\mu\nu}$. 

In comparison to a naive discretization method, where chromoelectric and -magnetic fields are replaced by lattice counterparts, our formulation improves the relative violation of the conservation laws by several orders of magnitude. The energy conservation equation \eqref{eq:firstContinuity} offers a means of obtaining the $T_{0k}$ components of the energy-momentum tensor that satisfies an exact energy conservation relation on the lattice. Challenges arise when deriving $T_{jk}$ using the momentum conservation equation \eqref{eq:secondContinuity}. The terms involving electric fields introduce spurious contributions like $E^jE^jB^k$ that are suppressed by the lattice spacing and are not present in the continuum expression. Furthermore, the terms involving magnetic fields cannot be written as a spatial derivative of $T_{jk}$ due to the lack of a suitable lattice Bianchi identity.
We have avoided this by replacing elements of $T_{jk}^B$ that are proportional to $\delta_{jk}$ with parts of the discrete energy density, which has led to a significant reduction of the violation. In the future, our focus will be on addressing these issues to obtain the subleading terms $\mathcal{O}(a_s^2)$ in the expressions of $T_{jk}$. 

For the energy-conserving continuity expression, we also see that the relative error due to finite timesteps can be further improved by orders of magnitude when using a time-symmetrized discretization of the equation. As illustrated by \eqs \eqref{eq:timeaverage1b} and \eqref{eq:timeaverage2}, the electric and magnetic fields then lie on the same time slice in the standard leapfrog algorithm. 

We expect our work to have several interesting applications, especially with an extension to account for small perturbations on top of a nonequilibrium plasma \cite{Kurkela:2016mhu,Boguslavski:2018beu,Boguslavski:2021buh}. Examples of these are transport coefficients, e.g., shear viscosity in an over-occupied gluon plasma. In this context, the energy conservation given by the continuity equation can hopefully be used to prevent the activation of other modes, like sound modes in the case of shear viscosity, ensuring a more accurate depiction of the system's behavior. 
We have made a first step toward a direct measurement of such transport coefficients in App.~\ref{app:EMT_pert}, where we have derived an expression for the perturbed energy-momentum tensor after introducing small fluctuations.

Another intriguing future research direction is to expand our work to a wider range of metric tensors, such as the Friedmann–Lema\^itre–Robertson–Walker (FLRW) metric and a longitudinally expanding (Bjorken) metric in the contexts of cosmology and heavy-ion collisions, respectively.
Indeed, including expansion in the framework would permit a more realistic treatment of the Glasma at the initial stages of heavy-ion collisions. Furthermore, it would extend the applicability of our framework to cosmological applications as well.

\section*{Acknowledgments} 
We would like to thank D.~I.~M\"uller, H.~Matsuda, and S.~Schlichting for valuable discussions. This work is supported by the European Research Council, ERC-2018-ADG-835105 YoctoLHC. This work was also supported by the European Union’s Horizon 2020 research and innovation by the STRONG-2020 project (grant agreement No. 824093). TL, JP, and PS have been supported by the Academy of Finland, by the Centre of Excellence in QuarkMatter (project 346324), and project 321840. KB would like to thank the Austrian Science Fund (FWF) for support under project P 34455. The authors wish to acknowledge the Vienna Scientific Cluster (VSC) under project 71444 and the CSC – IT Center for Science Finland, for computational resources on the supercomputer Puhti.

\appendix

\section{Derivation of a relation equivalent to the Bianchi Identity}
\label{app:Bianchi}

The general proof of eq.~\eqref{eq:ToProove} can be given as
\begin{align}
    \frac{\delta_{ij}}{2}[D_j,F_{mn}]F_{mn} &{}= \frac{\delta_{ij}\delta_{mm'}}{2}[D_j,F_{mn}]F_{m'n}\notag\\
    &{} = \frac{1}{2}\big(\epsilon^{kmj}\epsilon^{km'i} + \delta_{mi}\delta_{m'j}\big) [D_j,F_{mn}]F_{m'n}\notag\\
    &{}=\frac{1}{2}\epsilon^{kmj}\epsilon^{km'i}[D_j,F_{mn}]F_{m'n} +\frac{1}{2}[D_j,F_{in}]F_{jn}
\end{align}
where the first term can be further modified as
\begin{align}
    \frac{1}{2}\epsilon^{kmj}\epsilon^{km'i}[D_j,F_{mn}]F_{m'n} &{} = -\frac{1}{2}\epsilon^{kmj}\epsilon^{km'i}\epsilon^{mnl}[D_j,B^l]F_{m'n}\notag\\
    &{}=\frac{1}{2}\epsilon^{mkj}\epsilon^{mnl}[D_j,B^l]F_{m'n}\epsilon^{km'i}\notag\\
    &{}=\frac{1}{2}\big(\delta_{kn}\delta_{jl}-\delta_{kl}\delta_{jn}\big)[D_j,B^l]F_{m'n}\epsilon^{km'i}\notag\\
    &{}= \frac{1}{2}[D_j,B^j]F_{m'n}\epsilon^{nm'i} - \frac{1}{2}[D_j,B^k]F_{m'j}\epsilon^{km'i}\notag\\
    &{}=\frac{1}{2}[D_j,F_{m'i}]F_{m'j}\notag\\
    &{}=\frac{1}{2}[D_j,F_{in}]F_{jn}
\end{align}
On combining the above two expressions, one recovers eq.~\eqref{eq:ToProove}.

\section{Energy momentum tensor for fluctuations}
\label{app:EMT_pert}

Over the years, significant effort has been devoted to investigating the linear response of non-Abelian plasma, focusing on the analysis of fluctuations superimposed on background fields~\cite{Kurkela:2016mhu, Boguslavski:2018beu, Boguslavski:2021buh}. Calculating the linear response involves decomposing the gauge field and electric field as follows:
\begin{align}
    A^i(x)\rightarrow A^i(x)+a^i(x)\\
    E^i(x)\rightarrow E^i(x)+e^i(x)
\end{align}
The link matrix representing the combination of background and fluctuating fields is expressed as
\begin{align}
     U_{i,m}^{BG + fluct} = \big(1+ ig a_{i,m}a^s_i\big)U_{i,m}
\end{align}
resulting in the following form of the plaquette:
\begin{align}
    U_{jk,m}^{BG + fluct}&{}=U_{jk,m} + \delta U_{jk,m}\notag\\
    &{}=U_{jk,m} + ig \Big( a_{j,m} a^s_j U_{jk,m} + U_{j,m}a_{k,m+j}a_k^sU^{\dagger}_{j,m}U_{jk,m} -U_{jk,m}U_{k,m}a_{j,m+k}a^s_jU^{\dagger}_{k,m} - U_{jk,m}a_{k,m}a^s_k\Big)
\end{align}
where $a_i^s$ is the lattice spacing along the $i$ direction. Following this 
decomposition, one can formulate expressions for the energy-momentum tensor, such as:
\begin{align}
    \delta T_{00,m}&{}= \sum_{j,k>0}\frac{-1}{g^2a_j^2a_k^2}\frac{1}{4}\Big(\rm {ReTr}\big(\delta U_{jk,m} + \delta U_{jk,m-j} + \delta U_{jk,m-k} + \delta U_{jk,m-j-k}\big)\Big)+\frac{1}{4}\sum_{j>0}2 e^{j,a}_mE^{j,a}_m + 2 e^{j,a}_{m-j}E^{j,a}_{m-j}
\end{align}
Similarly, the expression for the density along the $k$-th component of linear momentum can be provided as:
\begin{align}
     \delta T_{0k,m}&{}=\sum_{j>0}\frac{1}{2g}\frac{a_ja_k}{a_j^2a_k^2}\Big\{\rm{ReTr}\big(ie^j_{m+k}U_{-kj,m+k} + iE^j_{m+k}\delta U_{-kj,m+k} + i e^j_mU_{jk,m}  + i E^j_m\delta U_{jk,m} \notag\\
    &{}+ ie^j_{m-j+k}U_{-kj,m-j+k} + iE^j_{m-j+k}\delta U_{-kj,m-j+k} + ie^j_{m-j}U_{jk,m-j} + iE^j_{m-j}\delta U_{jk,m-j}\big)\Big\}
\end{align}

\bibliographystyle{JHEP-2modlong}
\bibliography{EMT}

\end{document}